
\input phyzzx
\Pubnum={$\caps UCD-93-8$}
\date{February, 1993}


\def\gev{~{\rm GeV}}
\def\tev{~{\rm TeV}}
\def\pbi{~{\rm pb}^{-1}}
\def\fbi{~{\rm fb}^{-1}}
\def\fb{~{\rm fb}}
\def\pb{~{\rm pb}}


\def\prdj#1{{\it Phys. Rev.} {\bf D{#1}}}
\def\npbj#1{{\it Nucl. Phys.} {\bf B{#1}}}
\def\prlj#1{{\it Phys. Rev. Lett.} {\bf {#1}}}
\def\plbj#1{{\it Phys. Lett.} {\bf B{#1}}}
\def\zpcj#1{{\it Z. Phys.} {\bf C{#1}}}

\def\prepj#1{{\it Phys. Rep.} {\bf {#1}}}

%


\def\wtilde{\widetilde}

\def\anti{\overline}
\def\etal{{\it et al.}}
\def\etc{{\it etc.}}
\def\ls#1{\ifmath{_{\lower1.5pt\hbox{$\scriptstyle #1$}}}}

%

    \def\fillboxx#1#2{\hbox to #1{\vbox to #2{\vfil}\hfil}
    }

\def\mt{m_t}
\def\mb{m_b}

\def\hsm{\phi^0}
\def\mhsm{m_{\hsm}}

\def\mh{m_h}
\def\hl{h^0}
\def\mhl{m_{\hl}}
\def\hh{H^0}
\def\mhh{m_{\hh}}
\def\hpm{H^{\pm}}
\def\mhpm{m_{\hpm}}

\def\saariselka{{\it Proceedings of the Workshop on ``$\epem$ Linear Colliders
at $500\gev$: the Physics Potential''},  publication DESY 92-123A (1992),
ed. P.M. Zerwas,
Feb. 4 - Sept. 3 (1991) --- Munich, Annecy, Hamburg, Saariselka}
\def\calcutta{{\it Proceedings of the `Workshop in High Energy
Physics Phenomenology --- II'}, Calcutta (January, 1991)}
\def\aachen{{\it Proceedings of the Large Hadron
Collider Workshop}, edited by G. Jarlskog and D. Rein, Aachen (1990),
CERN 90-10, ECFA 90-133}

\def\smv{{\it Proceedings of the 1990 DPF Summer Study on
High Energy Physics: ``Research Directions for the Decade''},
editor E. Berger, Snowmass (1990)}

\def\ifmath#1{\relax\ifmmode #1\else $#1$\fi}
\def\half{\ifmath{{\textstyle{1 \over 2}}}}

\def\3quarter{{\textstyle{3 \over 4}}}

\def\tauptaum{\tau^+\tau^-}
\def\mupmum{\mu^+\mu^-}

\def\h{h}
\def\mh{m_{\h}}
\def\lplm{l^+l^-}
\def\eps{\epsilon}
\def\tanb{\tan\beta}

\def\mw{m_W}
\def\mz{m_Z}
\def\hpm{H^{\pm}}
\def\mhpm{m_{\hpm}}
\def\mhp{m_{\hp}}
\def\hp{H^+}
\def\hm{H^-}

\def\rta{\rightarrow}

\def\sq{\widetilde q}

\def\gl{\widetilde g}
\def\chitil{\widetilde\chi}

\def\cnone{\chitil^0_1}

\def\wp{W^+}
\def\wm{W^-}

\def\mstone{M_{\tilde t_1}}
\def\msttwo{M_{\tilde t_2}}

\def\msq{M_{\tilde q}}

\def\hl{h^0}
\def\hh{H^0}
\def\ha{A^0}
\def\mhl{m_{\hl}}
\def\mhh{m_{\hh}}
\def\mha{m_{\ha}}
\def\eps{\epsilon}

\def\lam{\lambda}

\def\gam{\gamma}

\def\gam{\gamma}

\def\VEV#1{\langle #1 \rangle}

\def\tanb{\tan\beta}

\def\chitil{\widetilde \chi}

\def\gl{\widetilde g}

\def\wp{W^+}
\def\wm{W^-}
\def\mw{m_W}
\def\mz{m_{Z}}

\def\lam{\lambda}

\def\hpm{H^{\pm}}

\def\hp{H^+}
\def\hm{H^-}

\def\mhpm{m_{\hpm}}

\def\epem{e^+e^-}
\def\mupmum{\mu ^+ \mu ^-}

\def\lplm{l^+ l^-}

\def\rta{\rightarrow}

\def\cnone{\chitil^0_1}

\def\mt{m_t}
\def\mhp{m_{H^+}}

\def\thetaw{\theta_W}

\def\M2pm{M^2_{P^\pm}}
\def\m2z{\mz^2}

\def\gexp{\Gamma_{\rm exp}}
\def\gmax{\Gamma_{\rm res}}
\def\h{h}
\def\mh{m_{\h}}
\def\yh{y_{\h}}
\def\gammah{\Gamma_{\h}}
\def\vevlam{\VEV{\lami\lamii}}
\def\eepem{E_{\epem}}
\def\egamgam{E_{\gam\gam}}
\def\lami{\lam}
\def\lamii{\lam^{\,\prime}}

\def\msusy{M_{SUSY}}

\def\anti{\overline}
\def\mstop{M_{\wtilde t}}


\def\wp{W^+}

\headline={\ifnum\pageno=1\firstheadline\else
\ifodd\pageno\rightheadline \else\leftheadline\fi\fi}
\def\firstheadline{\hfil}
\def\rightheadline{\hfil}
\def\leftheadline{\hfil}
        \footline={\ifnum\pageno=1\firstfootline\else\otherfootline\fi}
\def\firstfootline{\rm\hss\folio\hss}
\def\otherfootline{\hfil}
\font\tenbf=cmbx10
\font\tenrm=cmr10
\font\tenit=cmti10
\font\elevenbf=cmbx10 scaled\magstep 1
 1
\font\elevenit=cmti10 scaled\magstep 1

\line{\hfil }
\vglue 1cm
\hsize=6.0truein
\vsize=8.5truein
\parindent=3pc
\baselineskip=10pt
\centerline{\tenbf DETECTING THE SUPERSYMMETRIC HIGGS BOSONS:}
\vglue 0.2cm
\centerline{\tenbf AN OVERVIEW AND THE ROLE OF $\gam\gam$ COLLISIONS}
\vglue 1.0cm
\centerline{\tenrm JOHN F. GUNION}
\baselineskip=13pt
\centerline{\tenit Davis Institute for High Energy Physics,
Department of Physics}
\baselineskip=12pt
\centerline{\tenit University of California at Davis, Davis, CA, 95616, U.S.A.}
\vglue 0.8cm
\centerline{\tenrm ABSTRACT}
\vglue 0.3cm
{\rightskip=3pc
 \leftskip=3pc
\Tenpoint
\baselineskip=12pt
 \noindent

I summarize the abilities of LEP-II, the SSC/LHC hadron colliders,
and a future high energy $\epem$ collider (NLC) to detect the Higgs bosons
of the Minimal Supersymmetric Model (MSSM).
In the case of LEP-II, I emphasize the importance of having
as large a value of $\sqrt s$ as possible in order to
guarantee discovery of the light CP-even MSSM Higgs boson ($\hl$)
regardless of the top quark and stop squark masses.
For the SSC/LHC particular emphasis
is placed upon the comparison between results
for large supersymmetric particle mass scales as opposed to
smaller mass scales for which radiative corrections to the MSSM
Higgs sector are smaller and neutralino-chargino pair decays
of the MSSM Higgs bosons are kinematically allowed.
I then focus on the unique capabilities
of a back-scattered laser beam facility (BSLBF) at
the NLC for detecting the heavier
neutral Higgs bosons (the $\hh$ and $\ha$) of the MSSM.
In particular, I emphasize that
typical GUT scenarios predict that the $\hh$ and $\ha$
are likely to be very difficult to detect at the SSC and LHC,
and that $\mhh$ and $\mha$ can easily be too large to allow their discovery at
a $\sqrt s\sim 500\gev$ NLC. In contrast, I show that a
BSLBF could well allow $\hh$ and $\ha$ discovery
via inclusive $\gam\gam\rta \hh,\ha$ production out to masses of order
$0.8\sqrt s$, \ie\ roughly $400\gev$, at such an NLC. In addition, a BSLBF
allows one to check for CP violation in the neutral Higgs sector; such
CP violation is predicted to be absent in the MSSM.

\vglue 0.8cm }
\line{\elevenbf 1. Introduction \hfil}
\smallskip
\baselineskip=14pt
\twelverm
Despite the enormous success of the Standard Model (SM) of
strong and electroweak interactions,
the mechanism responsible for mass generation/electroweak symmetry
breaking (EWSB) remains uncertain. The naturalness and hierarchy
problems of the minimal Standard Model Higgs sector mechanism are
well-known.  Fully consistent models employing the general concept
of technicolor/compositeness remain elusive, and phenomenological
constraints on technicolor from precision LEP data are very restrictive.
Currently, supersymmetric models appear to be substantially more attractive.
First, supersymmetry
\REF\haberkane{For a general review and useful appendices on the minimal
supersymmetric model see H.E. Haber and G.L. Kane, \prepj{117C} (1985) 75.}
\refmark\haberkane\
is the only known theoretical framework that resolves the naturalness/hierarchy
problems while maintaining the elementarity of Higgs bosons.
\REF\hhg{J.F. Gunion, H.E. Haber, G. Kane and S. Dawson,
{\it The Higgs Hunter's Guide}, Addison-Wesley, Redwood City, CA (1990).}
(A detailed review of Higgs bosons in supersymmetric models is contained
in Ref.~[\hhg].) Further, most supersymmetric
models can be consistently incorporated into a simple grand unification
scenario while maintaining perfect agreement with the LEP data.
Indeed, the very simplest Minimal Supersymmetric Model (MSSM)
is completely successful in all these respects, even requiring no
intermediate mass scale(s) for grand unification.
\REF\gut{U. Amaldi \etal, \prdj{36} (1987) 1385;
G. Costa, J. Ellis, G.L. Fogli, D.V. Nanopoulos and F. Zwirner,
\npbj{297} (1988) 244; J. Ellis, S. Kelley and D.V. Nanopoulos,
\plbj{249} (1990) 442 and {\bf B260)} (1991) 131;
S. Kelley, J.L. Lopez, D.V. Nanopoulos, H. Pois and K. Yuan,
\plbj{273} (1991) 423;
P. Langacker and M. Luo, \prdj{44} (1991) 817;
U. Amaldi, W. de Boer and H. Furstenau, \plbj{260} (1991) 447;
I. Antoniadis, J. Ellis, R. Lacaze and D.V. Nanopoulos,
\plbj{268} (1991) 188;
I. Antoniadis, J. Ellis, S. Kelley and D.V. Nanopoulos,
\plbj{272} (1991) 31;
H. Aranson, D. Castano, B. Keszthelyi, S. Mikaelian,
E. Piard, P. Ramond and B. Wright, \prlj{67} (1991) 2933.}
\REF\rossroberts{G. Ross and R. Roberts, \npbj{377} (1992) 571.}
\REF\zichichi{F. Anselmo, L. Cifarelli, A. Petermann and A. Zichichi,
{\it Il Nuovo Cimento} {\bf 104A} (1991) 1817 and {\bf 105A} (1992) 581;
preprints  CERN/PPE-91-123, CERN/PPE-92-103,CERN-TH-6429-92, CERN-TH-6543-92.
F. Anselmo, L. Cifarelli and A. Zichichi,
preprints CERN/PPE-92-122, CERN/PPE-92-145.}
\refmark{\gut,\rossroberts,\zichichi}

Because of its consistency and simplicity, it is not unreasonable
to consider the MSSM as a real candidate theory appropriate for
describing physics between the electroweak scale and the grand unification
scale.  Thus, it is important to determine whether the experimental
information required to either rule out or confirm the model can be obtained
at present and future accelerators.
Experimental signatures for supersymmetry, deriving from
the many superpartner particles (such as the gluino, squarks, neutralinos
and charginos) are plentiful, both at $\epem$ and hadronic supercolliders.
However, the MSSM in its most general form has many parameters.
Consequently, even if all of these superpartners are discovered
a full test of the MSSM will be impossible without detecting its
Higgs bosons. Indeed, only by verifying the Higgs structure
predicted by the model can we be certain that the theoretical
difficulties associated with EWSB in the SM
are fully resolved in the manner predicted by minimal
supersymmetry. Further, the Higgs sector of the MSSM is so highly
constrained that even rather limited information about the Higgs
sector could rule out the model, and at the very least require
consideration of non-minimal supersymmetric models.  Or conversely,
verification of a few of the very specific predictions for the
Higgs sector could provide the first convincing evidence in favor of
the MSSM. Much effort has gone into developing techniques for detecting
the Higgs bosons of the MSSM in recent years.  I will review the
substantial progress that has been made and discuss the possibly
crucial role that could be played by a back-scattered laser beam
facility (BSLBF) at a future $\epem$ linear collider.

\vglue 0.6cm
\line{\elevenbf 2. Scenarios \hfil}
\vglue 0.4cm

Before proceeding with our detailed discussion, it is perhaps useful
to elaborate upon some of the possible experimental scenarios
that could arise. First, let us recall\refmark\hhg\
that the MSSM contains exactly two Higgs doublets and therefore predicts
the existence of five physical Higgs bosons, the CP-even
\foot{Classification of the neutral Higgs bosons by CP properties is
possible in the MSSM because the Higgs sector is automatically CP
conserving.}
$\hl$ and $\hh$ (with $\mhl<\mhh$), the CP-odd $\ha$, and a charged pair
$\hpm$.
It also requires that there be a gluino ($\gl$), squark partners for
all the quarks ($\sq\,\,$'s), two charginos ($\chitil^+_{1,2}$),
and four neutralinos ($\chitil^0_{1,2,3,4}$) with $\cnone$ being
the lightest supersymmetric particle. At tree-level,
the Higgs sector is completely determined by two basic parameters,
normally taken to be $\tanb=v_2/v_1$ (the ratio of the vacuum
expectation value for the neutral member
of the Higgs doublet, $\phi_2$, coupling to up quarks
to that of the doublet, $\phi_1$, coupling to down quarks)
and the mass of the CP-odd scalar, $\mha$. However,
the masses and couplings of the Higgs bosons
are strongly influenced by one-loop radiative corrections that
depend upon the top quark mass, $\mt$, and the masses of the
top squarks, $\mstone$ and $\msttwo$ (and, to a much weaker extent,
on other MSSM parameters). A complete review of radiative corrections was given
in earlier talks and can also be found in
\REF\perhaber{H.E. Haber, preprint UCSC-92/31, to appear in
{\it Perspectives in Higgs Physics}, ed. G. Kane,
World Scientific Publishing (1992).}
Ref.~[\perhaber].
\REF\pergunion{J.F. Gunion, preprint UCD-92-20, to appear in
{\it Perspectives in Higgs Physics}, ed. G. Kane,
World Scientific Publishing (1992).}
The results for Higgs masses, couplings, and branching ratios have
already been discussed in the earlier talks and
only the most important points will be repeated here.
Reviews can be found in Refs.~[\perhaber] and [\pergunion].
Taking $\mstone=\msttwo\equiv\mstop$,
the most crucial result obtained after radiative corrections is that,
for fixed $\mha$ and $\tanb$, the mass of the lightest CP-even Higgs boson
increases significantly with $\mt$ and $\mstop/\mt$
(the most important term in the one-loop radiative correction to $\mhl$
is proportional to $\mt^4\ln (\mstop/\mt)$). For given values of $\mstop$
and $\mt$, the maximum value of $\mhl$ occurs for large $\mha$ and large
$\tanb$. For large $\mha$, the $\hl$ has couplings that
approach those of the SM Higgs boson and the $\hh$, $\ha$ and $\hp$
all become rather degenerate in mass. It is in this region of
parameter space that a back-scattered laser beam facility is likely
to play a most crucial role.

Particularly important unknowns in determining the
ability of LEP-II, the NLC and the SSC/LHC to discover the Higgs bosons
of the MSSM relevance are:
the mass  scale for the superpartner particles; and, the
Higgs sector parameters $\mha$ and $\tanb$.  (The top quark will presumably be
discovered before long, and $\mt$ will be known.)
One scenario is illustrated by taking GUT schemes seriously.
Typically (see, for example, Ref.~[\rossroberts])
correct predictions for $\mw$ and $\mt$ at low energy,
starting from simple boundary conditions at the GUT scale $M_X$,
can only be obtained without fine-tuning if the
soft supersymmetry breaking mass scales are modest in size.
This would imply that the gluino, the squarks, and the neutralinos and
charginos are all relatively light.  In addition, the $\hl$ typically
has a mass $\gsim 100\gev$ (after radiative corrections)
and $\mhh\sim\mhpm\sim\mha\gsim250\gev$. (If certain $M_X$ boundary
conditions are relaxed, the superpartner particles would remain
relatively light, but the $\hh$, $\ha$ and $\hpm$ masses could be even
larger.) In such a case, LEP-II, unless it has energy somewhat above
$200\gev$, would not be able to detect any of these new particles.
At a new $\epem$ collider with energy of order $500\gev$ (dubbed the NLC),
the $\hl$ would be discovered, and the neutralinos and charginos could be
studied. But, in the scenario being considered the $ZZ\hh$
and $Z\hl\ha$ couplings are suppressed, implying that the only production
mode for the $\hh$ and $\ha$ with good coupling strength would be
$Z^*\rta \hh\ha$. However, this mode would not be kinematically allowed.
Similarly, the main mode for charged Higgs detection,
$Z^*\rta\hp\hm$ pair production, would not be kinematically allowed.
At the NLC, the only possibility for detecting any of the
Higgs bosons other than the $\hl$ would be to produce the $\hh$ and $\ha$
singly using back-scattered laser beams. As described later,
prospects for their detection in this manner are reasonable for
$\mha$ and $\mhh$ up to $\sim 300-400\gev$ (depending on $\tanb$).

At the SSC or LHC, the above scenario would yield a plentiful array of
signals for the gluino and the squarks, and large numbers of all of the
Higgs bosons would be produced.  The $\hl$ would again
most probably be detectable (if not already found at LEP-II);
it would be quite similar in properties to a SM Higgs boson of
the same mass.  However, detection of both the $\ha$
and $\hh$ would be very difficult since they would tend to decay
to two-body neutralino-neutralino or chargino-chargino channels,
which would have large backgrounds and missing energy.  Meanwhile,
the $\hp$ would be too heavy to appear in $t\rta \hp b$ decays
(the only case for which it has been clearly demonstrated that
detection of the $\hp$ would be possible). Indeed, the $\hp$ would
decay to a combination of $t\anti b$ and chargino-neutralino
channels, both of which have large backgrounds.

Thus, if the typical mass scale of the superpartner particles
(denoted generically by $\msusy$) is relatively low,
it could happen that many of the superpartner particles could be studied
at the NLC and/or SSC/LHC, but that only the SM-like $\hl$ of the Higgs
sector could be found.

However, without the restriction on fine-tuning, the GUT
schemes are all in agreement that $\msusy$ could be
large.  Although the original motivation for SUSY based on naturalness and
hierarchy considerations would begin to be lost if $\msusy$ were
so large that the superpartner particles could not be found at the
NLC or SSC/LHC, we certainly should allow for the possibility that
their masses (in particular those of the squarks and the gluino)
could be of order 1 TeV. In such a scenario, the first direct
indications and the most definitive tests of the MSSM could arise via its Higgs
sector. As for the previous scenario, the $\hl$ would be found, if not
at LEP-II then certainly at the NLC and most
probably at the SSC/LHC. However, detection of the $\ha$, $\hh$ and $\hp$
would remain quite problematic at the SSC or LHC, while
at the NLC the largest mass reach would be afforded by a BSLBF.

With these scenarios in mind, it is perhaps useful to give a few
explicit examples of how even very limited knowledge of the Higgs sector
could either rule out the MSSM or provide strong evidence in its favor.
The theory and phenomenology behind these examples has been given
in earlier talks and is reviewed in Ref.~[\pergunion].
Disproofs of the MSSM which would be possible if only the $\hl$ were
detected or excluded include the following observations.
\item{}
Failure to detect the $\hl$ with $\mhl\leq 160\gev$ (assuming
$\mt<200\gev$ and $\mstone,\msttwo<2\tev$); LEP-II with energy
$>240\gev$, or the NLC and/or SSC/LHC would be needed to fully
cover this $\mhl$ range.
\item{}
Detection of an $\hl$ with a value for $\mhl$ that is too small or
otherwise inconsistent with the observed value of $\mt$ and established
lower bounds on $\mstone,\msttwo$ and $\mha$.
\item{}
Detection of an $\hl$ with
$\mhl$ larger than the maximum allowed for large $\mha$ and large $\tanb$
and observed values for $\mstone$, $\msttwo$ and $\mt$; determination of the
stop squark masses would require their detection at SSC/LHC.
\item{}
Detection of an $\hl$ with
inconsistent values for or ratios between its coupling constants;
\eg\ if the $\hl ZZ$ coupling is SM-like, the $\hl t\anti t$
coupling should also have a SM-like value in the MSSM.

\noindent
If more than one Higgs boson is detected,
additional possible means for ruling out the MSSM arise.
These include:
\item{}
Detection of the $\hl$ and $\ha$ with
$\mha>2\mz$ but couplings for the $\hl$ that are not SM-like.
\item{}
Detection of the $\hl$ and $\ha$ with masses
$\mhl$ and $\mha$ that are inconsistent with
the observed $\mt$ and $\mstone,\msttwo$
values for all $\tanb$ choices.
\item{}
Two or more of the $\ha$, $\hh$ and $\hpm$ found with masses that are large
but not approximately degenerate.

\noindent The following observations would also contradict the MSSM.
\item{}
Detection of a neutral Higgs boson with mass larger than $2\mw$
{\it and} a large width associated with $WW/ZZ$ decays.
\item{}
Detection of CP violation in the Higgs sector.

\noindent
The general point should now be clear.  The Higgs sector of the MSSM
is so constrained that there are many precise relations and predictions
that can be checked, especially once $\mt$ and the masses of the stop
squarks have been measured (so as to determine the
most important radiative corrections to the Higgs boson masses.)
Of course, if results for any of the preceding consistency
checks were positive, this would constitute considerable support for the MSSM.
Certainly,  the relative simplicity and cleanliness of such tests
should provide more than enough motivation for finding means
for detecting the MSSM Higgs bosons for as large a range of parameter
space as possible.

\vglue 0.6cm
\line{\elevenbf 3. Masses, Coupling Constants and Branching Ratios \hfil}
\vglue 0.4cm

It will be useful to recall some of the more important results.
As already noted, at tree level all the
masses and couplings of the Higgs bosons are
determined by just two parameters, conventionally chosen to be
$\mha$ and $\tanb=v_2/v_1$.
All other Higgs boson masses as well as the neutral sector mixing
angle, $\alpha$, and all couplings to quarks and vector bosons can be expressed
in terms of these parameters.  In particular, one finds
$\mhl<{\rm Min}\{\mz,\mha\}|\cos 2\beta|$,
$\mhh\geq \mz$, and $\mhpm\geq \mw$.
Further, for large $\mha$, $\mhh\simeq\mha\simeq\mhpm$.
However, at one loop, additional parameters are required to fully
determine the masses and couplings of all the Higgs bosons.
Aside from $\mha$ and $\tanb$, values for
$\mt$ and a number of supersymmetric model parameters
(squark masses, $\mu$ and the $A_{b,t}$) must be specified.
In much of what follows, we shall simplify the discussion by neglecting
squark mixing and assuming the squarks to be degenerate.
In this case, the only important parameter other than $\mt$
for determining radiative corrections to the Higgs sector
is the common squark mass, typified by that of the stop squark, $\mstop$.

Perhaps the most important impact of the radiative corrections
is that as $\mt$ increases so does
the upper bound on $\mhl$; and, at the same time, the lower bound
on $\mhh$ gets larger.
The largest $\mhl$ values are attained in the large $\tanb$,
large $\mha$ corner of parameter space.  For $\mt=150\gev$ the largest
value is slightly in excess of $108\gev$, while for $\mt=200\gev$
the upper limit on $\mhl$ is about $138\gev$.
\REF\hemphaber{H.E. Haber and R. Hempfling, preprint SCIPP-91/33 (1991).}
\foot{
These results, and those of later plots, are obtained using
the complete leading-log radiative corrections to the MSSM Higgs
masses and couplings given in Ref.~[\hemphaber].}
Meanwhile, in the large
$\tanb$, small $\mha$ corner of parameter space are found the minimum
$\mhh$ values, $\sim 110\gev$ ($\sim 141\gev$) for $\mt=150\gev$ ($\mt
=200\gev$).  For $\mt=100\gev$ the upper bound on $\mhl$ and lower bound
on $\mhh$ are both near $\mz$, as predicted at tree level.
Note also that at large $\mha$ the approximate degeneracy
$\mha\simeq\mhh\simeq\mhpm$ continues to hold after including radiative
corrections.

Next, let us recall that some of the most crucial couplings of the
MSSM Higgs bosons are directly determined by $\cos^2(\beta-\alpha)$.
A remarkable feature of the MSSM is that $\cos^2(\beta-\alpha)$
decreases very rapidly with increasing $\mha$, and in fact is highly suppressed
over all of parameter space, except in the moderate to
large $\tanb$, small $\mha$ corner.
These features were first observed at tree-level,\refmark\hhg\ and
continue to pertain after radiative corrections, although radiative
corrections do decrease the suppression somewhat at every $\tanb$, $\mha$
parameter choice.
\REF\hhzz{R. Bork, J.F. Gunion, H.E. Haber, A. Seiden, \prdj{46} (1992) 2040.}
\REF\azz{J.F. Gunion, H.E. Haber, and C. Kao, \prdj{46} (1992) 2907.}
(More details can be found in, for instance, Refs.~[\hhzz-\azz].)
The importance of $\cos^2(\beta-\alpha)$ becomes apparent by
recalling the pattern of the couplings of the $\hl$
and $\hh$ to $VV$ ($V=Z$ or $W$) and $Z\ha$.
The squares of these couplings at tree-level are given by:
$$
\eqalign{\hl VV:&~~ f_V^2\sin^2(\beta-\alpha)\cr
         \hh VV:&~~ f_V^2\cos^2(\beta-\alpha)\cr}
\qquad
\eqalign{\hl \ha Z:&~~ {g^2\over 4 \cos^2\theta_W} \cos^2(\beta-\alpha)\cr
         \hh \ha Z:&~~ {g^2\over 4 \cos^2\theta_W} \sin^2(\beta-\alpha)\,,\cr}
\eqn\couppattern
$$
where $f_W=g\mw$ and $f_Z=g\mz/\cos\theta_W$.
(Of course, the $\ha$ has no $VV$ coupling at tree-level.)
Note the complementarity of the $ZZ$ and $\ha Z$ couplings
for the $\hl$ and $\hh$ --- in particular, if $\mha$ is large,
so that $\cos^2(\beta-\alpha)$ is small, the $\hl$ has large coupling to $ZZ$
while the $\hh$ has large coupling to $\ha Z$. If the angle
$\alpha$ is computed by diagonalizing the {\it radiatively corrected}
CP-even sector mass matrix, these results for the couplings remain valid
to order $g^4$. The one-loop corrections to
the vertices themselves first generate corrections to these
\REF\hhcoupnew{D. Pierce and A. Papadopoulos, \prdj{47} (1992) 222.}
\REF\pokhhzz{P.H. Chankowski, S. Pokorski, and J. Rosiek,
\plbj{286} (1992) 307.}
squared couplings at order $g^6$.\refmark{\azz,\hhcoupnew}\
We have not included the latter corrections
in the numerical work to be discussed later; they would not alter the basic
phenomenology.

Regarding quark couplings, the $t\anti t$  ($b\anti b$)
couplings of the $\hh$ and $\ha$
tend to be suppressed (enhanced)
at large $\tanb$, while those of the $\hl$ become SM-like at large
$\mha$. The most important radiative corrections to these couplings
are automatically included by using their tree-level forms as a function
of $\alpha$ and $\beta$, but with $\alpha$ determined
by diagonalizing the radiatively corrected CP-even mass matrix.
The coupling of the charged Higgs boson to $t\anti b$ is such that
the potentially large $\mt$ term is suppressed for large $\tanb$.

We note here one important impact of the the suppression of the
$WW,ZZ$ couplings of the $\hh$ and $\ha$.
This suppression implies that the $VV$ decays of the $\hh$ and $\ha$
generally have quite small partial widths.  This, in turn, implies
that even when heavy these Higgs
bosons are generally quite narrow resonances ($\Gamma<5\gev$).
The only situations
for which the $\hh$ width becomes large are two: if $\tanb\lsim 1$,
implying a large $t\anti t$ coupling for the $\hh$,
and $\mhh$ is above $t\anti t$ threshold; or if $\tanb> 20$,
in which case the $b\anti b$ coupling and, hence,
decay width of the $\hh$ is very enhanced. The behavior of
the $\ha$ width is very similar.  Of course, the $\hl$,
which is strongly coupled to $VV$ channels, cannot be heavy enough
(for $\mt\lsim 200\gev$ and $\mstop\lsim 2\tev$) to decay to $VV$
pairs, and thus will also be a narrow resonance.

Should the $\hh$ or $\ha$ be discovered, their
narrow width will be immediately apparent.  (Indeed,
the SSC/LHC detection techniques for these Higgs bosons rely on their having
a narrow width.) If, in addition, the observed state has
mass above $2\mw$, it will be obvious that it
cannot be the SM Higgs boson; the latter is expected
to rapidly develop a broad width in $VV$ decay channels for $\mhsm>2\mw$.
Conversely, discovery of a neutral
scalar with a large decay width to $VV$ channels would immediately
contradict the predictions of the MSSM.

The branching ratios clearly play a crucial role in understanding
how to detect the Higgs bosons of the MSSM.
It is a straightforward, if tedious exercise
to compute the branching ratios.  This has been carried out by
a number of groups, with consistent results.
\REF\bargersusy{V. Barger, M.S. Berger, A.L. Stange, and R.J.N. Phillips,
\prdj{45} (1992) 4128.}
\REF\baeretal{H. Baer, M. Bisset, C. Kao and X. Tata,
\prdj{46} (1992) 1067.}
\REF\kzsecond{Z. Kunszt and F. Zwirner, \npbj{385} (1992) 3.}
\REF\yamada{A. Yamada, preprint YITP-K-951-REV (1992),
{\it Mod. Phys. Lett.} {\bf A7} (1992) 2877.}
\refmark{\hhzz,\azz,\bargersusy-\yamada}\
Typical results were illustrated in earlier talks.
We first summarize crucial features of these results assuming that
neutralino and chargino pair decay modes of the Higgs bosons are forbidden.

\FIG\brshhino{}
\pageinsert
\vbox{
\phantom{0}\vskip 4.00in
\phantom{0}
\vskip .5in
\hskip -40pt
\special{ insert scr:perspectives_brs_hh_mt150.ps}
\vskip .5in
\phantom{0}\vskip 2.5in
\phantom{0}
\vskip .6in
\hskip -40pt
\special{ insert scr:erice_92_brsino_hh_mt150.ps}
\vskip -1.65in }
{\rightskip=3pc
 \leftskip=3pc
 \Tenpoint\baselineskip=12pt
\noindent Figure~\brshhino:
We plot branching ratios for $\hh$ decays for $\tanb=2$ and $20$
(at $\mt=150$) as a function of $\mhh$.
We have taken $\mstop=1\tev$ and neglected squark mixing.
In the top half of the figure, ino masses are large and ino-pair channels
are forbidden.  In the bottom graphs,
ino masses are determined by $M=200\gev$ and $\mu=100\gev$.
In the latter case only the more important decay channels are shown.
}

\endinsert

For $\mhl$ near its maximum value, it has SM-like couplings.
This means that it can have a significant $\gam\gam$ branching ratio,
and that the $ZZ^*$ branching ratio can become significant if
$\mhl^{max}$ is large enough (which requires large $\mt$ and $\mstop$).
Note that the $\tauptaum$ branching ratio is typically of order 10\%.

Branching ratios (in the case of large SUSY mass scales)
for the $\hh$ are illustrated in the top half of Fig.~\brshhino.
Note that the $ZZ^*$ and $ZZ$ branching ratios
are not necessarily suppressed despite the smaller-than-SM
width for decays to these channels.
If $\mhh<2\mt$ and $\tanb$ is not too large,
$\Gamma(\hh\rta ZZ)>\Gamma(\hh\rta b\anti b)$ and the $ZZ$ branching ratio
is substantial.  This means that the $\lplm\lplm$ detection mode
at the SSC/LHC could be viable.  A particularly interesting decay
of the $\hh$ is $\hh\rta\hl\hl$; this mode has a large branching ratio
for small to moderate $\tanb$.
Of course, for large $\tanb$ the $b\anti b$ and $\tauptaum$ modes
are dominant, with the latter branching ratio being of order 10\%.
For moderate to large $\tanb$, there is a narrow window of $\mhh$
for which $BR(\hh\rta\gam\gam)$ is reasonable.

The general pattern for the $\ha$ is similar, except for the severe
suppression of $VV$ decays (which are not present at tree-level,
first occurring at one-loop level). In addition,
$b\anti b$ and $\tauptaum$ branching fractions at $\tanb=2$
are larger than in the case of the $\hh$.
Also of interest is the $\ha\rta Z\hl$ decay that can have a large
branching ratio for moderate $\tanb$ if $\mha<2\mt$.  The $\gam\gam$ branching
ratio of the $\ha$ is only significant near its peak value at $\mha= 2\mt$,
unless $\tanb\lsim 1$.

Turning to the $\hp$, we will only be concerned with the
situation where $\mt>\mhp+\mb$, and $\hp\rta t\anti b$
decays are forbidden. Except in a small
window of parameter space where $\hp\rta \wp\hl$ decays are allowed
and would dominate, the primary decay of the $\hp$ for $\tanb\gsim 1.5$ will
be to $\tau^+\nu_\tau$ ($BR$ near 100\% for $\tanb\gsim 2.5$).
This mode might play a crucial role at the SSC/LHC.
\REF\recenthp{R.M. Barnett, R. Cruz, J.F. Gunion, and B. Hubbard,
preprint UCD-92-18 (1992), to be published in \prdj{}.}
For plots see Ref.~[\recenthp].
Of course, in the alternative case where $\hp\rta t\anti b$ decays are allowed,
they will be dominant.

Of course, if ino-pair channels are kinematically allowed they can be quite
important.  This is because the Higgs bosons have
couplings to charginos and neutralinos that are proportional to $g$,
rather than $gm/(2\mw)$ (as would be typical of a quark or lepton of mass $m$).
Indeed, when allowed, the ino-pair channels can easily dominate the decays of
a Higgs boson.  To illustrate this point, I compare the $\hh$
branching ratios when ino decay channels are forbidden to those found when
the ino's are relatively light, determined by taking
the $SU(2)$ gaugino mass $M$ to be $200\gev$
(corresponding in the standard GUT approach to $m_{\wtilde g}\sim 600\gev$)
and the higgsino mixing parameter $\mu$ to be $100\gev$.
(For this choice of parameters the neutralino masses at $\tanb=2$
are 39, 103, 119 and 245 GeV, while the chargino masses are 61 and 243 GeV.)
Results for $\mt=150\gev$ and $\tanb=2$ and 20
are illustrated in Fig.~\brshhino.
Note that the $\chitil^0$ pair modes dominate below $2\mt$.
The $\hl\hl$ mode branching ratio remains large, but the $ZZ$ branching ratio
is substantially suppressed in comparison to the case where ino decays
are forbidden.  At large $\tanb$, ino-pair modes continue to be significant,
but the $b\anti b$ mode is still dominant.  The primary result is
that $\hh$ detection at the SSC or LHC in the $\hh\rta ZZ\rta 4l$ mode
becomes essentially impossible, whereas without ino decays this mode
would be useful for moderate to small $\tanb$ when $\mhh\lsim 2\mt$.

The impact of ino decays on the $\ha$ branching ratios is quite similar
in nature to that found in the case of the $\hh$.  The $\hl$ decays
are also affected; the $b\anti b$ is often dominated by
ino-pair modes for $\hl$ masses such that the latter are kinematically allowed.
Even the $\hp$ decays are significantly altered when $\tanb\lsim 2$;
for such $\tanb$, the $\hp\rta \tau^+\nu_\tau$ width (proportional
to the square of $\mt\tanb$) is not especially enhanced and
neutralino-chargino pair channels can dominate.

Clearly, allowing for decays of the MSSM Higgs bosons to ino-pair
channels can significantly reduce their detectability in the
even R-parity channels involving SM particles and other Higgs bosons.
This fact will have its largest impact at the SSC/LHC, where the
only thoroughly established detection techniques rely on these modes.
At an $\epem$ collider, we expect that all decay channels, including
the odd R-parity channels, can be used to isolate a Higgs signal
using relatively efficient cuts to reduce backgrounds.  Thus, we will
outline the detectability of the MSSM Higgs bosons at $\epem$ colliders
primarily on the basis of their production rates.

\vglue 0.6cm
\line{\elevenbf 4. The MSSM Higgs Bosons at $\bf\epem$ Colliders \hfil}
\vglue 0.4cm

\bigskip
\line{\elevenit 4.1. LEP and LEP-II \hfil}
\smallskip

In the context of the MSSM, the decays
$Z\rta Z^*\hl$ and $Z\rta\hl\ha$ will take place if the couplings
are adequate and they are kinematically allowed.
Failure to detect these signals at LEP with the integrated
luminosity accumulated to date
indicates that $\mhl\gsim 40$~GeV, $\mha\gsim 20\gev$, $\mhpm\gsim 40\gev$;
\REF\pdg{Particle Data Group, \prdj{45}, Part II (1992).}
no restriction on $\tanb$ is obtained. \refmark\pdg\ These lower bounds
will be pushed to near $\mz$ after LEP-II completes its
experimental search for $\epem\rta Z^\ast\rta\hl\ha$ or $Z\hl$.
In particular, an $\hl$ with $\mhl\sim\mz$ can be seen
at LEP-II provided that $\sqrt s\gsim200\gev$, $L\sim 500\pbi$ can
be achieved, and that efficient $b$-tagging is possible.
\REF\btaggun{J.F. Gunion and L. Roszkowski, preprint UCD-90-26
(October, 1990), \smv, p. 203.}
\REF\btagbarger{V. Barger and K. Whisnant, \prdj{43} (1991) 1443.}
\refmark{\btaggun,\btagbarger}
Unfortunately, after radiative corrections
$\mhl$ could be larger than $\mz$ if $\mt$ and $\mstop$ are large.

The impact of radiative corrections on the potential of LEP-II
\REF\barbfrig{R. Barbieri and M. Frigeni, \plbj{258} (1991) 395.}
\REF\erz{J. Ellis, G. Ridolfi, and F. Zwirner, \plbj{262} (1991) 477.}
for detection of the neutral Higgs bosons of the MSSM
was first considered in Refs.~[\barbfrig] and [\erz].
Consider, for example, $\sqrt s=200\gev$ and $L=500\pbi$.
If one assumes an overall detection efficiency of 25\% and
that 25 observed events are needed for discovery, then at $\mt=150\gev$
there are already substantial portions of $\mha$--$\tanb$ parameter space
(at moderate to large $\mha$ and $\tanb$) for
which discovery of the $\hl$ is not possible using either
$\hl Z$ or $\hl\ha$ associated production.
The precise region is sensitive to the integrated
luminosity and analysis efficiencies.
\foot{For $\mt=150\gev$ and $\sqrt s=200\gev$,
the region is also very sensitive to the precise value of $\mt$.
This is because $\mz+\mhl^{max}$ is fairly close to $\sqrt s$, and
small changes in $\mt$ cause significant shifts in $\mhl$.}
For instance, if 10 observed events are sufficient
or $L$ a factor of 2.5 larger, $\hl$ detection
would be possible at LEP-200 for $\mt=150\gev$ in all but the large
$\tanb$, large $\mha$ corner of parameter space.
For $\mt=200\gev$, the boundary line
for $\hl$ detection is rather insensitive to precise experimental
efficiencies;  the $\hl$ is simply too heavy to be produced
in the $\mha\gsim140\gev$ and $\tanb\gsim 2$ portion
of parameter space.

\FIG\stoptopcontours{}

\midinsert
\vbox{\phantom{0}\vskip 5in
\phantom{0}
\vskip .5in
\hskip -110pt
\special{ insert user$1:[jfgucd.rcsusyhiggs]erice_92_contour_lepii.ps}
\vskip -.15in }
{\rightskip=3pc
 \leftskip=3pc
 \Tenpoint\baselineskip=12pt
\noindent Figure~\stoptopcontours:
The discovery boundaries for the $\hl$ at LEP-II in $\mt$--$\mstop$
parameter space assuming that $\mha$ is large ($\mha=1\tev$ was used)
for a variety of $(\sqrt s,\tanb)$ values. We assume that
$L=500\pbi$ and require 100 total $\hl Z$ events (before efficiencies).
We have neglected squark mixing. The region where discovery is possible
lies to the left and below the boundary curves.}

\endinsert

Of course, the ability of LEP-II to detect the $\hl$ is also strongly
dependent upon the actual $\sqrt s$ that can be achieved and upon
the unknown value of $\mstop$ (and, to a lesser degree, the parameters
that control squark mixing).
\REF\bargertau{V. Barger, K. Cheung, R.J.N. Phillips and
A. Stange, preprint MAD/PH/704 (1992).}
\REF\gunorrstudy{J.F. Gunion and L.H. Orr, unpublished.}
This issue is studied in Refs.~[\bargertau] and [\gunorrstudy].
In Fig.~\stoptopcontours\
the contours for 100 total $\hl Z$ events (no efficiencies included)
are given for large $\mha$ ($\mha\gg\mz$)
in $\mt$--$\mstop$ parameter space for a selection of
$(\sqrt s,\tanb)$ values, assuming $L=500\pbi$.
Once $\mhl$ is larger than about $100\gev$ this number of events
should prove marginally adequate to observe the Higgs signal above
the background coming from continuum
$\epem\rta ZZ$ production.\refmark\gunorrstudy\
100 $\hl Z$ events would, for instance, yield $\sim 7$
events in which the $Z$ decays to $\lplm$ ($l=e,\mu$) and
the $\hl$ would be seen as a bump in the $(p_{e^+}+p_{e^-}-p_Z)^2$ spectrum.
\REF\janot{P. Janot, preprint LAL 92-27 (1992) and references therein.}
For most $(\sqrt s,\tanb)$ choices this criterion turns out to be
quite close to the criteria specified in the detailed
experimental treatment of Ref.~[\janot].  (The only exception is
for $(190\gev,20)$, where the criteria of Ref.~[\janot] yield a significantly
more pessimistic contour than that plotted in Fig.~\stoptopcontours.)
Of course, for $\mhl$ near to $\mz$,
$b$ tagging would be needed to eliminate the $ZZ$ continuum;
$\tau$ pair final states would also help to isolate the Higgs boson.
(See, for instance, Refs.~[\bargertau] and [\janot].)

The figure makes it quite apparent that the $\sqrt s=175\gev$ option for LEP-II
(more or less the minimum that is anticipated) is far less desirable
than the somewhat more expensive $\sqrt s= 200\gev$ possibility.
Pushing to the technically feasible $\sqrt s =240\gev$ energy
would be highly desirable since the $\hl$ would be
detectable for all but the very largest top and stop masses.
Of course, it can be hoped that the top quark will be found at
the Fermilab Tevatron in the near future.  This would provide crucial
guidance in setting the maximum LEP-II machine energy for which we
should strive.  For instance, if it is found that $\mt\simeq 140\gev$, then
Fig.~\stoptopcontours\ shows that discovery of the $\hl$ at LEP-II
would be possible for all of the most interesting region
of parameter space ($\mstop\lsim 1\tev$ and $\tanb\lsim 20$) for a machine
energy of $\sqrt s=200\gev$. In contrast, for this value of $\mt$
even a small decrease
in energy to $\sqrt s=190\gev$ would greatly worsen the prospects;
detection of the $\hl$ would not be possible for high
values of $\tanb$ if $\mstop\gsim 350\gev$. Even a lower bound placed
on $\mt$ by the Tevatron would be helpful. For example, if it is determined
that $\mt>160\gev$, the most reasonable procedure might be
to first push LEP-II to an energy of $200\gev$, thereby
allowing a significant possibility for $\hl$ detection
according to Fig.~\stoptopcontours. But, if the $\hl$ were not discovered
at this energy, pushing slowly to $\sqrt s=240\gev$ would rapidly
open up the possibility for $\hl$ detection in
the regions of parameter space corresponding to the higher values
of $\mt$, $\mstop$ and $\tanb$.

\bigskip
\line{\elevenit 4.2 The Next Linear $\epem$ Collider\hfil}
\smallskip

As we have just seen, if the MSSM model is correct, the $\hl$
may or may not be discovered at LEP-II, depending upon $\mt$, the precise
machine parameters, in particular $\sqrt s$, and the MSSM model
parameters, especially $\mstop$.
In the event that the $\hl$ is not observed, because
$\mstop$ is large and/or the $\sqrt s$ achieved is small,
a higher energy $\epem$ collider would clearly be required in
order to detect any of the MSSM Higgs bosons.  Even if the $\hl$
is found at LEP-II, as discussed earlier the $\hl$ will have
couplings to quarks and vector bosons that are quite similar
to those of a SM Higgs boson of the same mass should it happen
that $\mha\gsim2\mz$. In this {\it a priori} rather likely case,
discrimination between the Higgs sectors of the SM and the MSSM requires
detection of one or more of the additional MSSM Higgs bosons, \ie\
the $\ha$, $\hh$ and/or $\hpm$.
Not surprisingly, an
$\epem$ collider of sufficient energy and luminosity would be an ideal
machine for this purpose.  An overview of this subject is given
in Ref.~[\hhg]. For an early tree-level
survey of relevant production mechanisms
\REF\earlystudy{J.F. Gunion, L. Roszkowski, A. Turski, H.E. Haber,
G. Gamberini, B. Kayser, S.F. Novaes, F. Olness, and J. Wudka,
\prdj{38} (1988) 3444.}
and strategies see Ref.~[\earlystudy].
These early surveys were updated to include radiative corrections
to masses and coupling constants
\REF\nlcstudy{A. Brignole, J. Ellis, J.F. Gunion, M. Guzzo,
F. Olness, G. Ridolfi, L. Roszkowski and F. Zwirner,
in \saariselka, p. 613.}
in Ref.~[\nlcstudy]. (See also,
\REF\janotstudy{P. Janot, in \saariselka, p. 107.}
Refs.~[\janotstudy] and [\yamada].)

\FIG\nlc{}

\midinsert
\vbox{\phantom{0}\vskip 5.5in
\phantom{0}
\vskip .5in
\hskip -120pt
\special{ insert user$1:[jfgucd.rcsusyhiggs]perspectives_nlc.ps}
\vskip -.15in }
{\rightskip=3pc
 \leftskip=3pc
 \Tenpoint\baselineskip=12pt
\noindent Figure~\nlc:
Event number contours for an NLC with $\sqrt s=500\gev$ and $L=10\fbi$.
We have taken $\mt=150\gev$, $\mstop=1\tev$ and have neglected
squark mixing. We display contours for 30 and 100 total events
(no detection efficiencies included) for
the six most important processes at an NLC: $Z^*\rta \hl,\hh+Z$,
$Z^*\rta \hl,\hh+\ha$, and $\epem\rta \nu\anti\nu\wp\wm\rta \hl,\hh+X$.}

\endinsert

Detection of the $\hpm$ is most easily discussed.  For $\sqrt s\gg\mz,2\mhpm$,
one finds (in units of the standard $R$ --- one unit of $R$
at $\sqrt s=500\gev$ corresponds to $0.347\pb$)
$$
{\sigma(\epem\rta\gamma^*,Z^*\rta \hp\hm)\over \sigma(\epem\rta\gamma^*\rta
\mupmum)}={1+4\sin^4\thetaw\over 8 \sin^4 2\thetaw}\simeq 0.308\,.
\eqn\sighphm
$$
Including effects of threshold suppression, and analyzing the
various different final state channels, it is found that
charged Higgs masses up to about $0.4\sqrt s$ will be detectable
with an integrated luminosity of $10^3$ inverse units of $R$.
\Ref\komamiya{S. Komamiya, \prdj{38} (1988) 2158.}
Of course, if $t\rta\hp b$ decays are allowed, $BR(t\rta\hp b)$
can be substantial and detection of the $\hp$ and $\hm$
in $\epem\rta t\anti t$ events could also be possible,
\REF\hhghphm{See Sec. 4.1 of Ref.~[\hhg], and references therein.}
\REF\dreesroy{M. Dress and D.P. Roy, \plbj{269} (1991) 155.}
\refmark{\hhghphm,\dreesroy}\
but the $\mhp$ values accessible in this way would probably
be smaller than $0.4\sqrt s$.
For higher $\mhp$, one is forced to consider
$\epem\rta t \anti b \hp$ production, via radiation of the $\hp$
off a virtual $b$ or $t$ quark.
\Ref\djkalzer{A. Djouadi, J. Kalinowski, and P.M. Zerwas,
\zpcj{54} (1992) 255.}\
This process could easily have a lower threshold
than $\hp\hm$ pair production.  However, the cross section
is not large; only for $\tanb\gsim 20$ does it exceed $2\fb$
even for the smallest allowed value of $\mhp$ ($\gsim\mw$).

As already described,
the dominant decay modes of the $\hp$ depend very much on what channels
are kinematically allowed. If $\hp\rta t\anti b$ decays are forbidden,
then the $\hp\rta \tau^+\nu$ mode is dominant for $\tanb\gsim 2$,
{\it unless} $\hp\rta\wp\hl$ is allowed.  When allowed, the $t\anti b$
decay mode will generally dominate.
\REF\hpbarnett{R.M. Barnett, I. Hinchliffe, J.F. Gunion, B. Hubbard,
H.E. Haber, and H.-J. Trost, \smv, p. 82.}
\REF\hpdecays{For more details see Refs.~[\hhg] and [\hpbarnett].}
None of these decay channels
should present any particular difficulty for detection of the $\hp$
at an $\epem$ machine provided the production rate is reasonable.

Let us now turn to the neutral Higgs bosons.
The discussion presented here will be largely based on the work
of Ref. [\nlcstudy].
The most important processes for production and
detection of the neutral Higgs bosons, $\hl$, $\hh$ and $\ha$, are six:
$\epem\rta Z^*\rta \hl,\hh+Z$, $\epem\rta Z^*\rta \hl,\hh+\ha$, and
$\epem\rta \nu\anti\nu+{\wp}^*{\wm}^*\rta \nu\anti\nu+\hl,\hh$
(the latter is conventionally referred to as $WW$ fusion).
There is considerable complementarity
among the first four processes, and the $WW$ fusion processes
are also complementary to one another and to the first four.
If $\mha\gsim \mz$, so that $\cos^2(\beta-\alpha)$ is small,
then $Z^*\rta \hl Z$, $Z^*\rta \hh\ha$, and
$WW\rta \hl$ are all maximal, and the other
three small.  The first three processes provide a reasonable event rate so
long as $\sqrt s$ is not too near the relevant threshold.
Typically, for $\sqrt s=500\gev$ and the $\mhl$ range of interest, $WW\rta\hl$
is comparable to $Z^*\rta \hl Z$, both having cross sections
$\gsim 0.1$ unit of $R$. Away from threshold,
the cross section for $Z^*\rta \hh\ha$
is at most about $0.1$ units of $R$.
(Precise cross section formulae appear in Ref.~[\hhg].)
Consequently, detection of the $\hh$ and $\ha$ will
only be possible in such a scenario if $\sqrt s$ is significantly
larger than $\mha+\mhh$ (\ie\ $\gsim 2\mha$ for large $\mha$
where $\mhh\simeq\mha$).
To illustrate more precisely the expectations, we give event number
contours for all six processes in Fig.~\nlc\ for an NLC with
$\sqrt s=500\gev$ and $L=10\fbi$ in the case where $\mt=150\gev$
and $\mstop=1\tev$ (squark mixing is neglected).  Our estimate
is that if more than 100 total events (before efficiencies)
are obtained for a given process, then it will be detectable.
Contours for 30 events are also shown, but detection of any of the processes
with so few events would require very high experimental and analysis
efficiencies.  The most important conclusion from this figure
is that detection of {\it all} of the neutral Higgs bosons will
be possible at $\sqrt s=500\gev$ if $\mha\lsim 200-220\gev$.
Detection of the $\hh$ and $\ha$ would require
higher machine energy for larger $\mha$ values.

While the above processes are certainly the most important ones
for detecting the MSSM Higgs bosons at the NLC, there are other
reactions that have been studied.  These include:
radiation of Higgs bosons off of top quarks or bottom quarks;\refmark\djkalzer\
$\hh$ and $\hl$ production via $ZZ$ fusion;\refmark\nlcstudy\
$\ha$, $\hh$ and $\hl$ production via the $\gam\gam$ collisions
arising from bremsstrahlung and beamsstrahlung radiation;
\refmark{\earlystudy,\nlcstudy}\ and production of pairs of Higgs
bosons or a vector boson plus a Higgs boson.\refmark{\earlystudy,\nlcstudy}
Generally speaking, the cross sections for these processes are substantially
smaller than those discussed earlier.  However, they could provide
important complementary information under some circumstances.

For example, Higgs radiation off of the top quark would probe
the $t\anti t~Higgs$ coupling that does not enter directly into the
production processes considered earlier.
At $\sqrt s=500\gev$, the $\hl t\anti t$ cross section is of order $1-3\fb$
for $\mhl$ near its upper limit (\ie\ for large $\mha$) where it has
SM--like couplings.  Detection of the process in several $10\fbi$
years might prove possible.  For small $\mha$, where $\mhl$ is
significantly below its maximum, the $\hl t\anti t$ cross section
declines rapidly.  The behavior of the $\hh t\anti t$ cross section
is complementary.  For small $\mha$, where $\mhh$ is near its lower limit,
and the $\hh$ couplings become SM--like,
$\sigma(\hh t\anti t)$ reaches the $1-2\fb$ range and a viable
signal might emerge.  As $\tanb$ increases, the $\hl$ ($\hh$) mass
regions for which they have SM--like couplings and
for which $\hl(\hh) t\anti t$ could be detected become
increasingly smaller. Finally, the $\ha t\anti t$ event rate
is never large enough for detection for $\tanb\gsim 1$.

Radiation of $\hl$, $\hh$ or $\ha$
off of a bottom quark could be useful at very large $\tanb$ where the
$b\anti b~Higgs$ couplings are highly enhanced.
For example, the $\ha t\anti t$ cross section is in the $1-3\fb$ range
for $\mha\lsim 120\gev$ if $\tanb\gsim 40$.  Unfortunately,
for larger $\mha$ values ($\gsim 200\gev$) such that $\ha,\hh$
discovery via $Z^*\rta\hh\ha$ fails, the $\ha t\anti t$ process
also becomes undetectable (even though phase-space-allowed)
unless $\tanb\gsim200$.

\vglue 0.6cm
\line{\elevenbf 5. Detecting MSSM Higgs Boson at Hadron Super Colliders\hfil}
\vglue 0.4cm

Detection of the MSSM Higgs bosons at a hadron super collider is
a much trickier subject, but a very important one since both the
SSC (with $\sqrt s=40\tev$) and the LHC (with $\sqrt s\sim 16\tev$)
will almost certainly be constructed prior to a high energy
linear $\epem$ collider.  Much work in this area has appeared
recently and has been reviewed in earlier talks.
A comprehensive review can also be found in Ref.~[\pergunion].
Overall, one can come
close to establishing a no-lose theorem according to which
one or more of the MSSM Higgs bosons will be detected either
at LEP-200 (if LEP-II only reaches $\sqrt s=175\gev$ the
theorem is greatly weakened) or at the SSC/LHC hadron super colliders.
To a significant extent, the same modes that have proven
critical to detecting the SM Higgs boson ($\hsm$) can be employed
for the MSSM Higgs bosons, provided that the decays of the MSSM
Higgs bosons are primarily to SM particle channels.
However, should the decays of a given MSSM Higgs bosons be dominated
by supersymmetric particle pair
channels, detection in the latter modes is likely to be necessary.
The degree to which this is possible is not currently known.
However, since a particularly important component of the no-lose theorem
involves discovery of the relatively light $\hl$, and supersymmetric
partners are less likely to dominate its decays,
we shall see that the no-lose theorem is not dramatically
weakened even when detection of the $\hh$ and $\ha$ in even R-parity
channels becomes impossible due to their having large
branching ratios to odd R-parity channels.

Before summarizing the situation, it is necessary to
briefly review the basic modes for discovery of a SM
Higgs boson with $\mz\lsim\mhsm\lsim 800\gev$ at the SSC/LHC.
\Ref\ochapters{For reviews of the details see, for example, Ref.~[\hhg] and
{\it Perspectives in Higgs Physics}, ed. G. Kane,
World Scientific Publishing (1992),
references therein, and other talks at this conference.}

\bigskip
\line{\elevenit 5.1. Techniques for Detecting
the SM Higgs Boson and Application to the MSSM\hfil}
\smallskip

Certainly one of the most difficult mass regions is
$80\lsim\mhsm\lsim135\gev$. In this region $\hsm\rta b\anti b$ decays
dominate, but have very large backgrounds.  Rare decay modes
have long appeared to provide the best hope.
\REF\gkw{J.F. Gunion, G.L. Kane and J. Wudka, \npbj{299} (1988) 231.}
\refmark{\gkw}\ A viable signal in inclusive
production, followed by $\hsm\rta\gam\gam$,
emerges only if very excellent $\gam\gam$ mass resolution
and $\gam-jet$ rejection is possible.
While the required mass resolution may be
achieved by some detectors, jet rejection
remains an issue. Further, not all detectors will
have the required resolution. Thus, it has been important to develop an
alternative approach to the $80\gev\lsim\mhsm\lsim 135\gev$ mass
region between the lower limit of the $ZZ^*\rta 4l$ channel (to be reviewed
shortly) and the approximate upper limit for $\hsm$ discovery at LEP-II.
The most promising (because of low backgrounds)
appears to be $W\hsm$ associated production,
followed by $W\rta l\nu$ and $\hsm\rta\gam\gam$.
The two processes leading to such a final state are:
$W^*\rta W\hsm$ production
\REF\kks{R. Kleiss, Z. Kunszt, and J. Stirling, \plbj{253} (1991) 269.}
\REF\mangano{M. Mangano, SDC Collaboration Note SSC-SDC-90-00113.}
\REF\overview{J.F. Gunion, G.L. Kane \etal, {\it Overview and Recent
Progress in Higgs Boson Physics at the SSC}, \smv, p. 59.}
\refmark{\kks,\mangano,\overview}\
and $gg\rta t\anti t\hsm$ production, with one of the $t$'s decaying to
the observed $W$.
\REF\gghttgunion{J.F. Gunion, \plbj{261} (1991) 510.}
\REF\gghttmarciano{W. Marciano and F. Paige, \prlj{66} (1991) 2433.}
\refmark{\gghttgunion,\gghttmarciano}\
At the SSC the latter has a much higher event rate than the former.
Backgrounds have been investigated and found to be sufficiently small for
reasonably good $\gam\gam$ mass resolution
and easily-achieved $\gam-jet$ discrimination factors.
One obtains a viable $\hsm$ signal at the SSC
throughout the $80\gev\lsim\mhsm\lsim 135\gev$ mass region for
the canonical $L=10\fbi$ of integrated luminosity. (At the LHC,
$L$ substantially above $10\fbi$ is required, but
the full $100\fbi$ enhanced luminosity is not necessary.)
Since $S/\sqrt B$ values are generally quite healthy for the SM Higgs boson,
the $l\gam\gam$ final state remains practical even for an MSSM Higgs
boson with $\gam\gam$ branching ratio somewhat below the SM-like value.

In practice, it is primarily the $\hl$ which might be detected in
$\gam\gam$ or $l\gam\gam$ final states.
As the $\hl$ couplings become SM-like, its $\gam\gam$ and
$l\gam\gam$ phenomenology
will become extremely similar to that of a $\hsm$ of the same mass.
Thus, it will mainly be a question of the region of parameter space
for which the $\hl$ has couplings that are close to SM values.
As we shall see, the $\ha$ and $\hh$ can also be detected in
the $\gam\gam$ and $l\gam\gam$ modes,
but only in very limited regions of MSSM parameter space.

In actual fact, the $l\gam\gam$ (and possibly $\gam\gam$)
mode continues to be useful for $\hsm$
discovery up to $\sim150\gev$.  However, above $\sim 135\gev$
(but $\lsim 2\mz$) an even cleaner mode becomes available as well,
namely $gg$ fusion production of the $\hsm$
followed by $\hsm\rta ZZ^*\rta \lplm\lplm$ ($4l$ for short, $l=e$ or $\mu$).
\refmark\gkw\
\REF\sdcloi{SDC Collaboration, Letter of Intent, SDC report, SDC-90-151
(1990) and the SDC Technical Design Report, SDC report, SDC-92-201
(April, 1992).}
(For a sample experimental study, see Ref.~[\sdcloi].)
With appropriate cuts on the $Z^*$ mass, the only background of any
significance at all is that from continuum $ZZ^*$ production, and
the level of this latter background is extremely tiny.  Detection of
the $\hsm$ in this mass region becomes purely a matter of event rate.

Once $\mhsm>2\mz$, $\hsm\rta ZZ$ double-on-shell decay becomes possible,
and it is well-known that detection of the $\hsm$ in the gold-plated
$4l$ mode is possible up to masses of order $600-800\gev$,
depending upon the strength of the $gg$ fusion mechanism as determined
by $\mt$.  For masses below this, the $ZZ$ continuum background
is significantly smaller than the Higgs resonance bump.

The $4l$ mode is clearly extremely useful for the MSSM Higgs bosons.
As we have already emphasized, the $\hl$ behaves very much like
a SM $\hsm$ of the same mass once $\mha$ is sufficiently
above $ \mz$ that its couplings become SM-like.  Thus, if $\mhl\gsim 135\gev$
and $\mha$ is relatively large, the $\hl$ will be detectable in the $4l$
mode.  In the case of the $\hh$, it might be mistakenly concluded that
the suppression of the $\hh ZZ$ coupling would make detection in
the $4l$ mode quite difficult.  However, two very important counter-acting
effects must be recognized. First, as noted earlier, the $\hh$ becomes
quite narrow. This means that one is looking for a very narrow resonance
above the $ZZ$ continuum background, in contrast to the SM case where
the $\hsm$ is a relatively broad resonance and a much larger mass interval
must be considered in computing the background. Second, even
if $\Gamma(\hh\rta ZZ)$ is suppressed, it can still be larger than
or, at least, comparable to $\Gamma(\hh\rta Q\anti Q)$
($\Gamma(\hh\rta\hl\hl)$ also plays an important role).
If this is the case, then the $\hh\rta 4l$ branching ratio is {\it not}
particularly suppressed.  In practice, this favorable
situation applies for $\mhh<2\mt$ so long as $\tanb$ is not so large
that $b\anti b$ decays of the $\hh$ become highly enhanced. However,
once $\hh\rta t\anti t$ decays are allowed, the $4l$ decays
have too small a branching ratio for this to be a useful channel.

\bigskip
\line{\elevenit 5.2. SSC/LHC MSSM Cross Sections \hfil}
\smallskip

Of course, an important ingredient in the detectability of the
MSSM Higgs boson is their production cross sections.
In the case of the $\hp$ we have already noted that only $gg\rta t\anti t$
production followed by $t\rta \hp b$ (or the charge conjugate)
will allow for $\hp$ detection.  The $t\anti t$ cross sections are well-known.
In the case of the neutral Higgs bosons,
aside from the $W^*\rta W+Higgs$ contribution to the $l\gam\gam$
final state mode, the only cross sections of importance are all
induced by $gg$ collisions: $gg\rta h$, $gg\rta b\anti b h$ and
$gg\rta t\anti t h$ ($h=\hl,\hh,\ha$).  Of these, the $t\anti t h$
process is only important for the $l\gam\gam$ final state mode.
As reviewed elsewhere, $WW$ fusion does not make a
significant contribution to the relevant production rates.

The $gg\rta b\anti b\hl$ process could possibly be of importance
for $\hl$ detection
in the inclusive $\tauptaum$ decay mode; at large $\tanb$ both it
and $gg\rta \hl$ are greatly enhanced with respect
to the corresponding SM Higgs cross sections when $\mhl$ is not near
its maximum. Note, however, that as $\mhl$ approaches its maximum value
the $\hl$ couplings become SM-like, and the $gg\rta b\anti b \hl$
cross section does not play a significant role in contributing to
$\hl$ detection in the $4l$ mode.

Clearly $gg\rta \hh$ is crucial for $\hh$ discovery
in the $4l$ mode, and, it turns out, so is $gg\rta b\anti b \hh$
(especially in the $ZZ^*\rta 4l$ mass region).
Indeed, for $\mhh$ of moderate size
the latter can become so enhanced at large $\tanb$ as to dominate
the $gg\rta \hh$ inclusive mechanism.
\REF\dicwil{D. Dicus and S. Willenbrock, \prdj{39} (1989) 751.}
In fact, for large enough $\tanb$
the $gg\rta b\anti b\hh$ cross section can be far larger than
that for any mechanism for producing a SM Higgs boson of similar mass.
\refmark\dicwil\
This last remark also applies to $gg\rta b\anti b \ha$.  Thus, at
large $\tanb$, it is possible that the very large cross sections
for $\hh$ and $\ha$ production, coupled with their $\lsim 10\%$ branching
ratios to $\tauptaum$, could allow their detection in the $\tauptaum$ final
state, despite the impossibility of using this mode for the SM $\hsm$.

\bigskip
\line{\elevenit 5.3. SSC/LHC MSSM Higgs Detection Phenomenology\hfil}
\smallskip

\REF\wheppii{J.F. Gunion,
`An Overview of, and Selected Signatures for, Higgs Physics at Colliders',
preprint UCD-91-9, in \calcutta.}
Some examination of the relevant issues at tree level appeared early on in
Refs. [\hhg,\overview,\wheppii] (see also references therein) and
related experimental studies for the SSC,
\REF\kzfirst{Z. Kunszt and F. Zwirner, in \aachen, Vol.II, p. 578.}
and in Ref. [\kzfirst] and related experimental studies for the LHC.
Recently, the phenomenology of the MSSM Higgs bosons at the LHC and SSC
has been re-examined after including radiative corrections.
\REF\gunorr{J.F. Gunion and L.H. Orr, \prdj{46} (1992) 2052.}
\REF\bargersusyii{V. Barger, K. Cheung, R.J.N. Phillips and A.L. Stange,
\prdj{46} (1992) 4914.}
Here, I will summarize the results obtained in Refs.~[\hhzz,\azz,\gunorr],
where the last-noted reference contains the basic surveys discussed below.
Overlapping work appears in Refs.~[\bargersusy-\kzsecond,\bargersusyii];
there is general agreement on the conclusions.

We shall consider $\mha$ and $\tanb$ as our fundamental Higgs sector
parameters,
but, as noted earlier, to determine one-loop leading-log radiative corrections
we must also specify $\mt$, the squark masses, and other parameters
that determine the amount of squark mixing.
Although LEP provides the lower bound of $\gsim20\gev$ for
$\mha$ noted earlier, there are currently no experimental
constraints on $\tanb$. On the basis of renormalization group arguments
it is generally expected that $1\lsim\tanb\lsim m_t/m_b$.
\refmark\hhg\ Thus, we have considered the range $0.5\leq \tanb\leq 20$.
In addition, we must specify the chargino and neutralino masses.
These particles would dominate Higgs decays and strongly affect
one-loop induced processes (such as $\gam\gam$ decays of neutral Higgs)
if sufficiently light.
In this subsection we shall take $\msq=1\tev$ and neglect
squark mixing. We shall also assume
that the ino masses are all greater than $200\gev$
(implying, in the minimal no-intermediate-scale GUT unification scheme,
a gluino mass somewhat in excess of 1 TeV).
This will be termed scenario (A).
\foot{In this next subsection we shall consider the impact of lowering
these two SUSY mass scales.}
In this case,
we can explore $\mha\lsim 400\gev$ without including the ino's in
the Higgs decays. In addition, as we noted earlier, if the charginos
are more massive than about $200\gev$, then they
essentially decouple from one-loop contributions to the $\gam\gam$
couplings of the neutral Higgs bosons. We also note that, for the squark
mass assumed, squark loop contributions to the $gg$ and $\gam\gam$ couplings of
the neutral Higgs bosons are small.

In the following we wish to determine the extent to which
one or more of the Higgs bosons of the MSSM can be detected
throughout all of $\mha$--$\tanb$ parameter space at either LEP-II
or the SSC/LHC hadron colliders. Let us list and label the
most relevant discovery channels.
For LEP-II we determine the parameter regions over which
the a) $Z^*\rta Z\hl$ and b) $Z^*\rta \hl\ha$ processes
should provide a viable signal --- our discovery criterion will be to require
25 events, assuming $\sqrt s=200\gev$, $L=500\pbi$ and an overall
detection efficiency of 25\% (\ie\ $0.2~\pb$ of cross section is
demanded).  For the SSC and LHC, we considered,
in Refs.~[\gunorr,\hhzz,\azz], only the cleanest and least controversial
detection modes for the MSSM Higgs bosons.
We adopted an integrated luminosity of $L=30\fbi$ ---
a reasonably achievable goal for both colliders. To recapitulate,
in the case of the neutral Higgs bosons the clean modes
are the same as employed for the SM $\hsm$: c),d) detection
of the $\hl,\hh\rta ZZ,ZZ^*\rta 4l$ decay modes;
and e),f) detection in the $W\hl X,W\hh X \rta l\gam\gam X$ final state.
In the case of the $ZZ^*\rta 4l$ mode, where backgrounds are negligible,
we required 15 events after a factor of $\eps=35\%$ reduction for
cuts and efficiencies. For the $ZZ\rta 4l$ and
the $l\gam\gam$ cases, we required $S/\sqrt B\geq 4$ after cuts and
efficiencies. (The same $\eps=35\%$ reduction as above
was used for the $4l$ mode;
a more sophisticated treatment is employed for the $l\gam\gam$ mode.
\refmark\gunorr)  Not listed above and not included
in the contour plots to presented later is the $W\ha X\rta l\gam\gam X$
detection mode for the $\ha$.
It would be of considerable use if $\tanb\lsim 1.5$.
In the case of the charged Higgs boson, the only detection mode
considered is g) $t\rta \hp b$ where the $\hpm$ is detected
via an excess of $\tau\nu$ events
over universality expectations. This is an excellent technique so long as
as $\tanb\gsim 0.5$ so that the branching ratio for $\hp\rta \tau^+\nu$
is not severely suppressed.
\foot{For values of $\tanb\lsim 1.5$, the $\hp\rta cs$ mode is also useful.
In terms of the graphical contours to be presented later, its impact
is relatively small. It is included in the $\hp$ detection contours
presented in Ref.~[\pergunion].}
Our specific criterion is derived from the detailed studies
\Ref\hpi{R.M. Barnett, J.F. Gunion, H.E. Haber, I. Hinchliffe, B. Hubbard,
and H.-J. Trost, SSC publication SDC-90-00141 (1990); R. M. Godbole
and D.P. Roy, \prdj{43}, 3640 (1991); L$^*$ Collaboration Letter
of Intent, publication SSCL-SR-1154 (1990); EMPACT Collaboration Letter of
Intent, publication SSCL-SR-1155 (1990).}\
that have shown that the high rate of $t\anti t$ production at the SSC
allows detection of
$t\rta \hp b$ decays unless $BR(t\rta \hp b)$ is quite small.  Typically,
a very significant effect in the $\hp\rta \tau^+ \nu$ decay mode
can be observed at the SSC for
$L=30\fbi$ so long as $BR(t\rta\hp b)\times BR(\hp\rta \tau^+\nu_\tau)
\gsim 0.003$, even if single $b$-tagging
is required to isolate the $t\anti t$ events of interest from background.
We employ this $BR$ lower bound as our discovery criterion.

As a final note, it should be remarked that if very excellent $\gam\gam$
mass resolution and $\gam-jet$ rejection can both be achieved,
inclusive production of the $\hl$, $\ha$ and $\hh$ followed by detection
in $\gam\gam$ decays will supplement the $l\gam\gam$ detection technique.
This was studied in Refs.~[\gunorr] and [\baeretal], where it was shown that
the regions of parameter space for which the inclusive $\gam\gam$ mode
might allow Higgs boson discovery are very similar to those where
the $l\gam\gam$ mode is viable.  The potential advantage of being
able to confirm a signal found in the $l\gam\gam$ using
the inclusive $\gam\gam$ channel, or vice versa, is apparent.

Before presenting a representative summary graph, some additional
discussion should prove useful. We first briefly recapitulate
the discovery abilities of LEP-II (with $\sqrt s=200\gev$ and
$L=500\pbi$). Consider first the $Z^*\rta Z\hl$ production channel.
Our earlier discussion showed that it
will be viable so long as it is kinematically allowed and one
is not in the small $\mha$, large $\tanb$ parameter space corner,
where the $\hl ZZ$ coupling is suppressed.
For $\mt=100\gev$ the $\hl$ never gets so heavy
that $Z \hl$ production is forbidden, and, therefore,
this mode is visible everywhere except in the above-noted
parameter space corner.
For $\mt=150\gev$, the $\hl$ becomes too heavy when $\mha\gsim 100\gev$
and $\tanb\gsim 7-10$ (depending on $\mha$).
For $\mt=200\gev$, the $\hl$ becomes too heavy over most of parameter
space except for moderate $\mha$ and $\tanb\lsim 3$. Consider next
the $Z^*\rta \hl\ha$ detection mode. It is essentially always viable
for parameter choices in the small $\mha$, large $\tanb$ corner,
whatever the value of $\mt$ ($\leq200\gev$).
This is simply because not only is the required coupling substantial,
but also the $\hl$ and $\ha$ both have small enough masses that the process is
well below kinematic threshold.
In actual fact, at $\mt=100\gev$, there is a very narrow region centered
about $\mha\sim 65\gev$ with $\tanb\gsim 6$ where neither the $\hl Z$
nor the $\hl\ha$ reaction satisfies the strict discovery criterion of 25
events for a 25\% detection efficiency.  However, in this region one
does obtain at least 15 events in one reaction or the other for $L=500\pbi$.
Obviously, this region would be filled in after a couple of years of
running.

We next give an overview of the expectations for the SSC/LHC
hadron collider detection modes. Consider first the $4l$ mode.
For $\mt\leq150\gev$, the $\hl$ is never sufficiently heavy
($\mhl\gsim 130\gev$ is required)
that it could (even with full strength coupling)
have an observable $4l$ decay rate. By $\mt=200$, if $\mha$
is large enough (roughly $\mha\gsim 150\gev$) the $\hl$
becomes heavy enough and has a sufficient fraction of the
full SM-like $ZZ$ coupling that its $ZZ^*\rta 4l$ event rate
exceeds the 15 event requirement.  The detectability of the
$\hh\rta 4l$ decays is equally sensitive to $\mt$. For $\mt=100\gev$,
there is essentially no mass range for which the suppressed
$\hh\rta ZZ \rta 4l$ decays are not swamped by $\hh\rta t\anti t$.
By $\mt=150\gev$, $\hh\rta 4l$ can be detected for $\mz\lsim\mha
\lsim2\mt$ so long as $\tanb$ is not so big that the $4l$
mode is overwhelmed by $\hh\rta b\anti b$ decays.  The upper
mass limit above is, of course, fixed by the $t\anti t$ threshold,
while the lower limit is determined by when $\mhh$ falls
too far below the $ZZ$ threshold (roughly $\mhh\lsim 130\gev$).
By $\mt=200\gev$, for small $\mha$ the $\hh$ is heavier than the critical
$130\gev$, and, in addition, has sufficiently
substantial $ZZ$ coupling that the $4l$ mode is visible. At larger
$\mha$ (but $\mhh\leq2\mt$ still),
this coupling becomes progressively more suppressed, and at higher
values of $\tanb$ the $4l$ decay is swamped by the $\hh\rta b\anti b$ decays.
Once $\mhh\geq 2\mt$, $t\anti t$ decays are dominant and the
$\hh\rta 4l$ decays cannot be seen for any $\tanb$.
Finally, we note that the region over which the $4l$ channel
can be seen is not very strongly dependent upon the resolution.

With regard to the $l\gam\gam$ mode, there is only the tiniest
region of parameter space (in the vicinity of $\mha\sim 80-90\gev$
and with $\tanb\gsim 4$) for which the $W$ loop (generally suppressed) and
quark loops combine to yield a $\hh\gam\gam$ coupling and, hence,
$BR(\hh\rta\gam\gam)$ that is large enough for the $\hh$ to be visible in this
channel once cross sections and backgrounds are taken into account.
In the case of the $\ha$, a significant $\gam\gam$ coupling
must derive from the top quark loop and is only possible if $\tanb< 1$
so that the $\ha t\anti t$ coupling is not suppressed.
Thus, the $l\gam\gam$ mode is generally viable for the $\ha$ if
$\tanb\lsim 1$ and $\mha<2\mt$.
In the case of the $\hl$, the $l\gam\gam$ mode is always
visible if the $\hl$ has sufficiently SM-like couplings
that its $\gam\gam$ branching ratio is similar to that of
a $\hsm$ of the same mass, and if it is sufficiently
heavy ($\mhl\gsim 80\gev$) that $\gam\gam$ decays are not suppressed
and backgrounds not too large.
These criteria are satisfied, more or less independently of $\tanb$,
so long as $\mha$ is large enough.  For $\mt=200\gev$, the radiative
corrections cause $\mhl$ to reach
the required mass region for smaller $\mha$ than for $\mt=150\gev$,
while at $\mt=100\gev$, $\mhl$ is large enough only at quite large $\mha$.
For instance, at $\mt=200\gev$,
as $\mhl$ approaches its maximum value
for large $\mha$ and the $\hl$ couplings become SM-like,
the event rate is several times
the minimum required in order to achieve a $4\sigma$ signal.  Thus,
a reduction in the $\hl$ rate
as its couplings move away from the SM-like limit with declining $\mha$
can be afforded.

\FIG\surveyssci{}
\midinsert
\vbox{\phantom{0}\vskip 5in
\phantom{0}
\vskip .5in
\hskip -107pt
\special{ insert user$1:[jfgucd.rcsusyhiggs]erice_92_surveya_ssc_mt150.ps}
\vskip -.25in }
{\rightskip=3pc
 \leftskip=3pc
 \Tenpoint\baselineskip=12pt
\noindent Figure~\surveyssci:
Discovery contours in $\mha$--$\tanb$
parameter space for the SSC with $L=30\fbi$ and LEP-200 with
$L=500\pbi$ for the reactions: a) $\epem\rta
\hl Z$ at LEP-200; b) $\epem\rta \hl\ha$ at LEP-200; c) $\hl\rta 4l$;
d) $\hh\rta 4l$; e) $W\hl X\rta l\gam\gam X$; f) $W\hh X\rta l\gam\gam X$;
g) $t\rta \hp b$.
The contour corresponding to a given reaction is labelled by
the letter assigned to the reaction above. In each case, the letter
appears on the side of the contour for which detection of the
particular reaction {\it is} possible.
We have taken $\mt=150\gev$, $\mstop=1\tev$ and neglected squark mixing.}

\endinsert

We are now in a position to present a sample summary graph.
In Fig.~\surveyssci\ we give discovery contours for LEP-200 and the SSC,
for the case of $\mt=150\gev$. Integrated luminosities for
LEP-200 and the SSC of $500\pbi$ and $30\fbi$, respectively,
are assumed. The figure displays the regions in $\mha$--$\tanb$
parameter space for which discovery of one of the MSSM Higgs bosons
using the reactions a)-g) listed earlier will be possible.
Discovery criteria are as motivated and stated earlier:
$\geq 25$ events (after including a detection
efficiency of $\eps=0.25$) for reactions a) or b) at LEP-200;
$S/\sqrt B\geq 4$ for Higgs masses above $2\mz$, or 15 events
(after efficiencies) for Higgs masses below $2\mz$,
for reactions c)-d); $S/\sqrt B\geq 4$ for reactions e)-f);
and $BR(t\rta \hp b)\times BR(\hp\rta\tau^+\nu_\tau)\geq 0.003$
for the charged Higgs detection mode g).
In the following, we briefly summarize the most important
conclusions that can be reached from this and similar figures for other
$\mt$ values.

Fig.~\surveyssci\ shows that in the case of $\mt=150\gev$ detection of one
or more of the MSSM Higgs bosons will be possible either at LEP-200,
or at the SSC, except in a window (indicated by the ? mark
on the figure) with $\mha\sim 120-150\gev$ and $\tanb\gsim 8-10$.
\foot{In the following discussion, parenthetical letters refer to the process
labels appearing on the figures.}
For large $\mha\gsim 200\gev$ and $\tanb\gsim 4$,
only the $\hl$ will be detectable. For this same $\mha$ range
and $\tanb$ between $\sim 4$ and $\sim 8$ the $\hl$ will be found at
both LEP-200 and the SSC. But once $\tanb\gsim 8$ only the SSC can
detect the $\hl$. Note that detection of the $\hl$
at the SSC is only possible in the $l\gam\gam$ final state channel (e) ---
$\mhl$ is never large enough for the $ZZ^*\rta 4l$ mode to
have significant branching ratio.
Once $\mha\lsim 150\gev$, detection of the $\hl$ will not be possible
at the SSC, but discovery at LEP-200 (a,b) will be possible over a significant
fraction of this portion of parameter space.  Detection of the $\hh$
in the $4l$ mode (d) at the SSC is confined to small to moderate $\tanb$ and
$\mha$ between $\sim 60$ and $\sim 300\gev$ (decreasing as $\tanb$ increases).
Detection of the $\hh$ in the $l\gam\gam$ mode (f) is confined to a very
narrow region of parameter space centered on $\mha\sim 60\gev$
with $\tanb\gsim 5$.
Detection of the $\hp$ in $t$ decays (g) at the SSC will be possible
if $\mha\lsim 115\gev$.

\FIG\surveysscii{}
\midinsert
\vbox{\phantom{0}\vskip 5in
\phantom{0}
\vskip .5in
\hskip -107pt
\special{ insert user$1:[jfgucd.rcsusyhiggs]erice_92_surveya_ssc_mt200.ps}
\vskip -.25in }
{\rightskip=3pc
 \leftskip=3pc
 \Tenpoint\baselineskip=12pt
\noindent Figure~\surveysscii:
As for Fig.~\surveyssci, but for $\mt=200\gev$.}

\endinsert

A similar graph for $\mt=200\gev$ appears in Fig.~\surveysscii.
It shows that for large $\mt$ at least one MSSM Higgs boson can be found
everywhere in $\mha$--$\tanb$ parameter space --- there would be no `gap'.
Indeed, the SSC alone can
discover at least one of the MSSM Higgs bosons throughout all of
parameter space (whereas the region of parameter space
that is covered by LEP-200 is relatively limited for such a large $\mt$).
In particular,
at the SSC the $\hh\rta 4l$ channel (d) becomes viable in all but the
large $\mha$, large $\tanb$ region of parameter space (where $b\anti b$ decays
of the $\hh$ suppress the $4l$ decays of interest). Meanwhile, the $\hl$
can be discovered in both the $4l$ (c) and $l\gam\gam$ (e) modes for all
$\mha\gsim 200$ and $\tanb\gsim 2$, thereby allowing determination
of both its $ZZ$ and its $t\anti t$ couplings. For $\mt=200\gev$ and
$\mha\lsim 150\gev$, the $\hl$ will not generally be detectable
at the SSC, but both the $\hh\rta 4l$ (d) and $t\rta \hp b$ (g) processes will
be observable.  Detection of the $\hh$ in the $l\gam\gam$ channel (f)
even becomes significant in a substantial wedge of parameter space
with $\mha$ between $\sim 65$ and $\sim 105\gev$ and $\tanb\gsim 2.5$.

In contrast to these two larger $\mt$ cases,
at $\mt=100\gev$ detection of the MSSM Higgs bosons at the SSC
would be confined to a much more limited portion of parameter space.
In this case, radiative corrections are small and $\mhl$ would
never be significantly larger than $\mz$; indeed, $\mhl$ would be substantially
smaller than $\mz$ in many regions of parameter space.
Thus, the $4l$ channel would not be viable for the $\hl$ anywhere in
parameter space. However, for $\mha\gsim 250\gev$ and $\tanb\gsim 2$
it can be demonstrated (see Ref.~[\gunorr]) that the $\hl$ could be found
via the $l\gam\gam$ mode, the $\hl$ mass being $\gsim 80\gev$.
At low $\mha$, $t\rta\hp b$ decays could be observed.  A small
region with $\mha\lsim 200\gev$ and low to moderate $\tanb$
where $\hh\rta 4l$ decays could be observed would remain.
Thus, at this small a $\mt$ LEP-200 would play a very prominent role.
As already discussed, it is virtually certain that the $\hl$
could be detected.  But, of course, the only other Higgs boson
of the MSSM that might be found at LEP-200 would be the $\ha$
were it to have small enough mass that $Z^*\rta \hl\ha$ is kinematically
allowed and $\cos^2(\beta-\alpha)$ is large.

The above plots and discussion have omitted detection of the $\ha$
at the SSC. As already noted, and described in detail in Ref.~[\gunorr],
the $\ha$ can be detected at the SSC in the $l\gam\gam$ (and,
possibly, $\gam\gam$) mode if $\mha\leq 2\mt$ and $\tanb\leq 1$.
Thus, its discovery in such modes
at the SSC would instantly place a strong upper limit on $\tanb$.

A more complete discussion of the above outlined results and the
corresponding results for the LHC can be found in Refs.~[\gunorr,\hhzz,\azz]
as well as the other earlier-referenced work by other authors.
The LHC results are easily summarized:
the LHC with $L=100\fbi$ gives very much the same
discovery contours as the SSC with $L=30\fbi$.

\bigskip
\line{\elevenit 5.4. Influence of Ino Decays and Effects of Low $\mstop$ \hfil}
\smallskip

Of course, the results obtained above will be altered if the masses
of supersymmetric partner particles are substantially lower than
we have been assuming.  First,
radiative corrections to the MSSM Higgs boson masses
and mixing angle $\alpha$ become
relatively small if $\mstop$ is substantially smaller than $1\tev$.
In addition, relatively light neutralinos and charginos  will become important
for decays of the Higgs bosons, as illustrated
earlier in Fig.~\brshhino.  These two effects
have a substantial impact on both LEP-200 and SSC phenomenology.
To illustrate, I have considered several alternative scenarios.
\item{(A)} First, we have the case discussed above, with $\mstop=1\tev$ and
neutralinos and charginos all heavier than $200\gev$.
\item{(B)} Second, I consider $\mstop=1\tev$
with neutralino and chargino masses set
(as in Fig.~\brshhino) by $M=200\gev$ and $\mu=100\gev$.
\item{(C)} Third, I take squark masses as determined using
the above $M$ and $\mu$ values, a
low-energy soft-SUSY-breaking mass of $m_{\wtilde Q}=m_{\wtilde U}
=m_{\wtilde D}=300\gev$,
and soft-SUSY-breaking ``$A$'' parameters of $A_t=A_b=50\gev$.
Here, $\wtilde Q$, $\wtilde U$ and $\wtilde D$
are the doublet, up-singlet and down-singlet squark states, respectively.
The $m_{\wtilde Q,\wtilde U,\wtilde D}$
are taken to be the same for all three generations.
$A$ parameters for the first two families are taken to be the same
as for the third family, but, in any case, have negligible effect.
Full ``$D$'' and ``$F$'' term mass contributions are included,
and mass matrix diagonalization performed.

\noindent
In both (B) and (C), all allowed decays to ino pairs
are incorporated (see Fig.~\brshhino\ and associated discussion).
(Squark pair channels are still above threshold for the $\mha$
range considered here.) Exact contributions to $\gam\gam$ and $gg$ couplings
coming from chargino and/or squark loops are included for both decays
and production. Work related to that discussed below
\REF\bkbtd{H. Baer, C. Kao, M. Bisset, X. Tata and D. Dicus,
preprint FSU-HEP-920724 (1992).}
has appeared in Ref.~[\bkbtd].

Of course, in most GUT scenarios $\mstop$ would probably not be as
large as $1\tev$ if the neutralinos and charginos are light.
Thus, scenario (B) might be regarded as somewhat adhoc; but,
it presents a useful point of comparison. On the other hand,
scenario (C) is not unrepresentative of results found
in the specific grand unification analyses mentioned earlier.
For the choice of $\mstop\sim 300\gev$,
the radiative corrections to the $\hl$ mass are quite
small, and Fig.~\stoptopcontours\ shows that
LEP-II (with $\sqrt s=200\gev$) could detect at least the $\hl$
of the MSSM. With regard to the SSC, we can crudely anticipate that
detection of the $\ha$, $\hh$ and $\hp$ will become more difficult.
With this in mind, let us turn to specific results.

\FIG\surveysscib{}
\midinsert
\vbox{\phantom{0}\vskip 5in
\phantom{0}
\vskip .5in
\hskip -107pt
\special{ insert user$1:[jfgucd.rcsusyhiggs]erice_92_surveyb_ssc_mt150.ps}
\vskip -.25in }
{\rightskip=3pc
 \leftskip=3pc
 \Tenpoint\baselineskip=12pt
\noindent Figure~\surveysscib:
Discovery contours in $\mha$--$\tanb$ parameter space for scenario (B).
Notation \etc\ as for Fig.~\surveyssci.}

\endinsert

\FIG\surveysscic{}
\midinsert
\vbox{\phantom{0}\vskip 5in
\phantom{0}
\vskip .5in
\hskip -107pt
\special{ insert user$1:[jfgucd.rcsusyhiggs]erice_92_surveyc_ssc_mt150.ps}
\vskip -.25in }
{\rightskip=3pc
 \leftskip=3pc
 \Tenpoint\baselineskip=12pt
\noindent Figure~\surveysscic:
Discovery contours in $\mha$--$\tanb$
parameter space as for Fig.~\surveysscib\ but in scenario (C).
}

\endinsert

Results for scenarios (B) and (C) are presented in Figs.~\surveysscib\
and \surveysscic, respectively.  The main effect in going from (A)
to (B) is obviously the decrease in the $\hh\rta 4l$ viability region.
This occurs due to the smaller size of $BR(\hh\rta ZZ)$ as depicted
in Fig.~\brshhino.  In going from (B) to (C) the $\hh\rta 4l$
region almost disappears.  This is because $BR(\hh\rta ZZ)$
has declined further --- partly due to the fact that
the $\hh$ is less massive for low $\mha$ values when $\mstop$
is small, and partly due to the fact that $\cos^2(\beta-\alpha)$
(which determines the $\hh ZZ$ coupling) decreases significantly
as the radiative corrections to $\alpha$ decline with decreasing
$\mstop$. (See Refs.~[\hhzz,\azz] for relevant graphs.)
Meanwhile, the $\hl\rta l\gam\gam$ (and, for good $\gam\gam$ resolution,
also the $\gam\gam$ channel) region remains relatively stable.
The exception is the $\tanb\sim 1$ region.  In going from (B) to
(C) the radiative corrections to $\mhl$ decrease to such an extent
that for $\tanb$ near 1
$\mhl$ falls below the $\sim 80\gev$ lower limit for which
these modes are viable.
The $t\rta\hp b$ detection mode also remains relatively stable
for $\tanb\gsim 4$ since $\mstop$ does not greatly influence the $\hp$
mass and the $\chitil^+\chitil^0$ decay modes do not greatly decrease
the $\hp\rta \tau^+\nu_\tau$ branching ratio.  However,
for $\tanb\lsim 4$, the $\tau\nu$ decay channel width declines to
a level such that the ino-pair modes do become important.
Consequently, the maximum $\mhp$ that can be detected in this way decreases
significantly, which is reflected in the shift of the $\hp$
discovery contour to smaller $\mha$ values.
Finally, as noted earlier, in going from (B) to (C) the coverage
of parameter space by LEP-200 becomes almost complete.

\FIG\surveyssciib{}
\midinsert
\vbox{\phantom{0}\vskip 5in
\phantom{0}
\vskip .5in
\hskip -107pt
\special{ insert user$1:[jfgucd.rcsusyhiggs]erice_92_surveyb_ssc_mt200.ps}
\vskip -.25in }
{\rightskip=3pc
 \leftskip=3pc
 \Tenpoint\baselineskip=12pt
\noindent Figure~\surveyssciib:
Discovery contours in $\mha$--$\tanb$
parameter space as for Fig.~\surveysscib\ but for $\mt=200\gev$.
}

\endinsert

\FIG\surveyssciic{}
\midinsert
\vbox{\phantom{0}\vskip 5in
\phantom{0}
\vskip .5in
\hskip -107pt
\special{ insert user$1:[jfgucd.rcsusyhiggs]erice_92_surveyc_ssc_mt200.ps}
\vskip -.25in }
{\rightskip=3pc
 \leftskip=3pc
 \Tenpoint\baselineskip=12pt
\noindent Figure~\surveyssciic:
Discovery contours in $\mha$--$\tanb$
parameter space as for Fig.~\surveysscic\ but for $\mt=200\gev$.
}

\endinsert

Changing $\mt$ does not greatly alter these trends.
To illustrate, we present in Figs.~\surveyssciib\ and \surveyssciic\
the contours for $\mt=200\gev$ in scenarios (B) and (C), respectively.
Consider first Fig.~\surveyssciib.
The region for which $\hh\rta 4l$ detection is possible expands
greatly compared to the corresponding $\mt=150\gev$ case,
Fig.~\surveysscib, but is still somewhat reduced compared
to scenario (A) at $\mt=200\gev$, Fig.~\surveysscii.
Compared to scenario (A) at $\mt=200\gev$,
the $\hl\rta 4l$ detection region is decreased ---
keeping $\mstop$ fixed leaves $\mhl$ and, hence, $\hl\rta ZZ^*$
phase space unchanged, but ino-pair channels for the $\hl$ become important.
Another significant difference between the $\mt=150\gev$ and $200\gev$
scenario (B) cases arises from the fact that the
larger $\mt$ value expands the region of viability for $t\rta\hp b$,
to $\mha\lsim 150\gev$ for all $\tanb$ values. Let us now turn
to the scenario (C) case, Fig.~\surveyssciic.
Comparing Fig.~\surveyssciic\ to Fig.~\surveysscii, we see that
the impact of taking both $\mstop$ and the ino masses to be small
instead of large is quite dramatic.
Indeed, Fig.~\surveyssciic\ is much more similar
to Fig.~\surveysscic\ (the $\mt=150\gev$, scenario (C) figure).
$\hl$ detection in the $4l$ mode has become impossible because
of the much smaller radiative correction addition to $\mhl$
when $\mstop$ is small. $\hh\rta 4l$ has become impossible
for most of parameter space, for exactly the same reasons
outlined for $\mt=150\gev$. The $l\gam\gam$ (and $\gam\gam$)
mode for the $\hl$ remains viable for $\mha\gsim 150\gev$,
except for $\tanb\sim 1$.

A particularly important point to note is the following:
regardless of $\mt$, the SSC will
have difficulty detecting the $\hh$ (and $\ha$) neutral MSSM Higgs bosons
over most of parameter space if all SUSY mass scales are small, unless
the neutralino-chargino-pair decay modes can be isolated from backgrounds.
The authors of Ref.~[\bkbtd] have performed a first examination
of such ino-pair modes and conclude that they could provide viable signals
in rather limited regions of parameter space. But, even in these
hopeful regions, the backgrounds are significant, especially those coming from
continuum ino-pair production processes. To fully assess the possibilities,
detailed detector-simulation studies of the ino-pair modes are needed.

\bigskip
\line{\elevenit 5.5. The No-Lose Theorem and Overview\hfil}
\smallskip

The above remarks and figures can be summarized by saying that the combination
of LEP-II (with $\sqrt s=200\gev$, $L=500\pbi$) and the SSC
(with $L=30\fbi$) comes close to providing a no-lose theorem:
at least one of the MSSM Higgs bosons will be discovered
in one of the very robust modes discussed at
one or the other machine for any choice of the basic
parameters $\mt$, $\mha$, and $\tanb$. In order for the
coverage of the various detection channels to be sufficiently
complete that one is equally close to a no-lose theorem at the
LHC, the full enhanced luminosity of $L=100\fbi$ will be required.

In scenario (A), with large SUSY masses, we have seen that
LEP-II and the SSC are highly complementary at moderate $\mt\sim 150\gev$.
In contrast, at small $\mt$ LEP-II plays the major role, while at large $\mt$
LEP-II covers only a small fraction
of parameter space whereas the SSC covers all of parameter space. Of course,
if $\mt=150\gev$, the no-lose theorem is not actually quite complete
for large $\mstop$ values.
Without considering additional channels, there is a gap
for $\mha\sim 125$ to $160\gev$ and $\tanb\gsim 10$ for which
neither LEP-II nor the SSC can detect any of the MSSM Higgs bosons.
A decrease in the LEP-II machine energy to a $\sqrt s$ value
significantly below $200\gev$ would increase the size of this gap.
(This gap may be covered, at least in part, by other SSC discovery
channels that we shall discuss below.)

The other natural extreme is that in which SUSY mass scales are all modest
in size, scenario (C). We have seen that, regardless of the value
of $\mt$, LEP-II will play the most crucial role.
Low $\mstop$ implies that radiative
corrections to $\mhl$ become small implying that LEP-II will be able to
detect the $Z\hl$ process throughout much of parameter space.
The SSC will provide complementary information on $\hl\rta\gam\gam$
for $\mha\gsim 150\gev$. In addition, the SSC will always be able to detect
$t\rta\hp b$ decays, if kinematically allowed. However, at the SSC
detection of the $\hh$ (or $\ha$) in the $4l$, $\gam\gam$ or $l\gam\gam$ modes
or of $\hl\rta 4l$ events will not be possible in most of parameter space.
As already noted, scenario (C) is not atypical of that arising
in a GUT study of the MSSM.  Indeed, a typical GUT result predicts not
only that the squark and ino masses are modest in size, but also
yields moderately large values for $\mha$ and $\tanb$.
Typically, these latter parameters
fall in the range $200-400\gev$ and $5-30$, respectively.
As already noted, at the SSC only $\hl$ detection in the $l\gam\gam$
(and/or $\gam\gam$) mode will be viable for such parameter choices.

\bigskip
\line{\elevenit 5.6. Alternative Detection Modes: A Brief Note \hfil}
\smallskip

While the production/detection modes considered above potentially
provide the cleanest signals for the MSSM Higgs bosons, there
are a variety of other modes that clearly deserve consideration.
Those with substantial promise in the case where
neutralino/chargino pair decays are forbidden were reviewed
in Ref.~[\pergunion].
Full evaluation of the viability of all the modes will require detailed
Monte Carlo assessment of mixed electroweak/QCD backgrounds.
However, there is clear hope that
the signatures discussed will provide at least weak signals
at the SSC for the neutral Higgs bosons over important regions
of parameter space.  In particular, $\hh$ and $\ha$ detection
in the $\tauptaum$ decay channel seems likely to be practicable
at high enough $\tanb$ ($\gsim 15?$) in the moderate $\mha$
domain where the no-lose theorem as outlined earlier is incomplete.
This is due to the large enhancement of the $gg\rta \hh,\ha+b\anti b$
cross sections at large $\tanb$.

\vglue 0.6cm
\line{\elevenbf 6. A Back-Scattered Laser Beam Facility at the NLC\hfil}
\vglue 0.4cm

So far, we have seen that LEP-II, an NLC with $\sqrt s\sim 500\gev$,
and the SSC/LHC hadron colliders are guaranteed to detect at least
the $\hl$ throughout all of parameter space.  However, detection
possibilities for the $\hh$, $\ha$ and $\hp$ are likely to be very
limited at the hadron colliders and even the NLC only has a
guaranteed reach in mass to about $0.4\sqrt s$.  The latter
is restricted primarily by virtue of the fact that
for large $\mha$ the only high rate reactions are $\hh\ha$ and $\hp\hm$
pair production.
An extremely interesting possibility for extending the $\mha$ and $\mhh$
mass reach of a high-luminosity NLC is to use collisions of
two photons produced by back-scattering laser beams off of the incoming
$e^+$ and $e^-$ beams at the NLC.
\REF\telnovi{H.F. Ginzburg, G.L. Kotkin, V.G. Serbo, and V.I. Telnov,
{\it Nucl. Inst. and Meth.} {\bf 205} (1983) 47.}
\REF\telnovii{H.F. Ginzburg, G.L, Kotkin, S.L. Panfil, V.G. Serbo,
and V.I. Telnov, {\it Nucl. Inst. and Meth.} {\bf 219} (1984) 5.}
\refmark{\telnovi,\telnovii}\
Because the $\ha$ and $\hh$ would be produced singly, the kinematical
limit placed on detectable $\mha,\mhh$ values is of order $\sqrt s$.
We shall see that there is a good possibility that in practice
one can probe masses as high as $0.8 \sqrt s$. In addition,
production of the $\hl$, $\ha$, and $\hh$ in this way automatically
provides a measure of their $\gam\gam$ couplings --- couplings
of great interest as a probe of new physics in the loop graphs
from which they arise.
Detection of the SM Higgs boson and of the neutral Higgs bosons of
the MSSM using back-scattered laser beam photons was first studied in
\REF\ghbslaser{J.F. Gunion and H.E. Haber, preprint UCD-90-25
(September, 1990), \smv, p. 206; and preprint UCD-92-22 (1992).}
Ref.~[\ghbslaser]. Monte Carlo results for a SM Higgs boson have appeared
\REF\bordenetal{D.L. Borden, D.A. Bauer, and D.O. Caldwell,
preprint SLAC-PUB-5715 (January 1992).}
in Ref.~[\bordenetal].
Here, we shall be summarizing the results of Ref.~[\ghbslaser].

\FIG\bslaserhh{}

\midinsert
\vbox{\phantom{0}\vskip 5.0in
\phantom{0}
\vskip .5in
\hskip -40pt
\special{ insert scr:erice_92_bslaserhh.ps}
\vskip -1.65in }
{\rightskip=3pc
 \leftskip=3pc
 \Tenpoint\baselineskip=12pt
\noindent Figure~\bslaserhh:
We plot rates for $\gam\gam\rta \hh\rta Q\anti Q$ compared
to the $\gam\gam\rta Q\anti Q$ background ($Q=b$ or $t$) for $\gam\gam$
collisions produced by back-scattered laser beams at a linear $\epem$ collider.
We have required an angular cutoff of $z=\cos\theta<0.5$ on the
outgoing $Q$'s in the center of mass frame.  An effective (see text)
photon-photon luminosity of $L_{eff}=20\fbi$ has been employed and it is
assumed that mass resolution in the $Q\anti Q$ channel will be
$\Gamma_{exp}\sim 5\gev$. Finally, for incoming electrons and laser beams
such that $2\lam_e\lam_\gamma\sim 1$ we
have used the estimate of $\vevlam\simeq0.8$,
obtained after convolution over the two back-scattered photons, 1 and 2.
The relatively background free $\hh\rta\hl\hl$ and $\hh\rta ZZ$
decay mode event rates (computed with $\gmax=\infty$) are also shown
(no $ZZ$ decay branching fraction included).
It should be noted that in the region $165\lsim\mhh\lsim 220\gev$,
the decay $\hh\rta\hl\hl$ is kinematically forbidden for the MSSM
parameters chosen.
Supersymmetric partners are assumed to be sufficiently heavy that
they do not influence the $\hh\gam\gam$ coupling or $\hh$ decays.
Radiative corrections to the MSSM Higgs sector are incorporated with
$\msusy=1\tev$ and squark mixing neglected. Results for $\mt=150\gev$
and $\mt=200\gev$ are displayed for both $\tanb=2$ and $\tanb=20$.}

\endinsert

\FIG\bslaserha{}

\midinsert
\vbox{\phantom{0}\vskip 5.0in
\phantom{0}
\vskip .5in
\hskip -40pt
\special{ insert scr:erice_92_bslaserha.ps}
\vskip -1.65in }
{\rightskip=3pc
 \leftskip=3pc
 \Tenpoint\baselineskip=12pt
\noindent Figure~\bslaserha:
We plot rates for $\gam\gam\rta \ha\rta Q\anti Q$ compared
to the $\gam\gam\rta Q\anti Q$ background ($Q=b$ or $t$)
with the same conventions as for Fig.~\bslaserhh.
The exotic $\ha\rta Z\hl$
decay mode event rate (computed with $\gmax=\infty$) is also shown.
}
\endinsert

\FIG\bslaserhl{}

\midinsert
\vbox{\phantom{0}\vskip 5.0in
\phantom{0}
\vskip .5in
\hskip -40pt
\special{ insert scr:erice_92_bslaserhl.ps}
\vskip -1.65in }
{\rightskip=3pc
 \leftskip=3pc
 \Tenpoint\baselineskip=12pt
\noindent Figure~\bslaserhl:
We plot rates for $\gam\gam\rta \hl\rta b\anti b$ compared
to the $\gam\gam\rta b\anti b$ background for the same conditions
as specified in Fig.~\bslaserhh.}

\endinsert

The number of Higgs ($h$, $h=\hl$, $\ha$ or $\hh$) events in a given channel
$X$
is given by
$$\eqalign{
N(\gam\gam\rta\h\rta X)=&{8\pi BR(\h\rta X)\over \eepem
 \mh^2} \tan^{-1}{\gmax\over
\gammah} \Gamma(\h\rta\gam\gam) \cr
&\times F(\yh) (1+\vevlam_{\yh})
{\cal L}_{\epem}\,,\cr}\eqn\nevents$$
where $\yh\equiv \mh/\eepem$.
In Eq.~\nevents, $F(y)$ is defined in terms of the differential $\gam\gam$
luminosity as $F(y)\equiv {\cal L}_{\epem}^{-1}(d{\cal L}_{\gam\gam}/dy)$, and
$\gmax\equiv {\rm max}\{\gexp,\gammah\}$;
$\vevlam_{\yh}$ is the average product of the polarizations of the
colliding photons evaluated at $\yh$. For $Q\anti Q$ channels
$\gexp$ is taken to be the mass resolution that can be achieved.
Since both the $\ha$ and $\hh$ are generally narrower
than a few GeV in the MSSM, optimal signal to background ratio is achieved
by taking $\Gamma_{exp}$ as small as possible.  We shall employ
a relatively optimistic value of $\gexp\simeq 5\gev$.
In the limit where $\gexp\gg\gammah$ the inverse tangent approaches
$\pi/2$, and $N$ is independent of $\gmax$.
In channels $X$ that are free of background, $\gmax=\infty$ is employed,
which also results in replacing the inverse tangent by $\pi/2$.

For the MSSM Higgs bosons a variety of
channels are important: these include $X=Q\anti Q$ for all the neutral
Higgs bosons; $X=\hl\hl$ and, possibly, $X=ZZ$ for the $\hh$;
and $X=Z\hl$ for the $\ha$.
(Recall that $VV$ decays are absent at tree-level for the $\ha$.)
The analysis performed to date also assumes
that two-body decays to supersymmetric
particle pair channels are absent, \ie\ that neutralinos and charginos are
heavier than about half the Higgs boson mass.
Studies\refmark{\telnovi,\telnovii,\bordenetal}
of $F(y)$, as obtained by convoluting the spectra of the colliding
back-scattered photons, indicate that $F\sim 1-2$
is not atypical. In detail, both $F(y)$ and $\vevlam$
depend on the helicity of the incoming electron $\lam_{e}$
and the polarization of the laser beam $\lam_\gam$ with which it collides,
and on the corresponding parameters for the positron beam.
If the incoming electron and associated laser beam
are polarized so that $2\lam_{e^-}\lam_\gam\sim 1$ (and similarly for
the positron and its associated laser beam) then
values of $\vevlam\gsim 0.8$ can be achieved for a substantial
range of $\yh$ values corresponding to
$m_h$ below about 70\% of $E_{\epem}$.\refmark\bordenetal\
For $m_h$ values above this it is best to use
$2\lam_e\lam_\gam\sim -1$ so that $F(y)$ peaks for $m_h$ near $E_{\epem}$.
In this case, for $m_h$ in the vicinity of $ E_{\epem}$ one finds that
$\vevlam_{\yh}$ can again be large, but is rapidly varying
and could also be small.

The background event rate for a $Q\anti Q$ final state is given by
$$N(\gam\gam\rta Q\overline Q)= {\gmax\over \eepem}
F(\yh) {\cal L}_{\epem} \sigma_{Q\overline Q}(\mh^2,z_0)\,.\eqn\nbkgnd$$
In Eq.~\nbkgnd\
$$
\eqalign{
 \sigma_{Q\overline Q}(s,z_0)=
{4\pi\alpha^2e_Q^4N_c\over s}
\Biggl\{&-\beta z_0\left[1+{(1-\beta^2)^2\over (1-\beta^2z_0^2)}\right]
+\half\left[ 3-\beta^4\right]\log {1+\beta z_0\over 1-\beta z_0}\cr
&+{\lami\lamii}\beta z_0\left[1+{2(1-\beta^2)\over 1-\beta^2z_0^2}
-{1\over \beta z_0}\log {1+\beta z_0\over 1-\beta z_0}\right]
\Biggr\}\,.\cr}\eqn\integratedbkgnd$$
where $\beta\equiv \sqrt{1-4m_Q^2/m_h^2}$  and $z_0$ is
the angular cutoff ($|\cos\theta_Q|\leq z_0$)
on the $Q$ direction in the $\gam\gam$ center of mass.
There are two important observations regarding $\sigma_{Q\anti Q}$.
First, it is greatly suppressed by an angular cutoff of
order $z_0=0.85$, especially for $m_h\gg2m_Q$.
(In contrast, the Higgs signal is isotropic.) Second,
it is easily seen that $\sigma_{Q\anti Q}\propto
[\lami\lamii-1]$ if $\beta\simeq 1$. Thus,
for large $E_{\epem}$ we can severely suppress the $b\anti b$ background
in the region $m_h<2m_t$ by
polarizing the incoming electron and laser beams to achieve large
$\vevlam_{\yh}$.
\foot{Gluon radiation can allow the $Q\anti Q$ background to evade
the $[\lami\lamii-1]$ suppression factor if the gluon
is hard and not collinear with either of the $Q$'s.  However, an experimental
limitation on such radiation can in principle preserve the background
suppression.
\Ref\khoze{We thank V.A. Khoze for discussions on this issue.}
}
Once $m_h>2m_t$ we must focus on the
$t\anti t$ channel for which both techniques are somewhat less
effective, but still of importance.

In assessing backgrounds to the $\hl\hl$ and $Z\hl$ channels, we assume
that the $\hl$ will already have been discovered and its mass fairly precisely
determined.  Its primary decay will be to $b\anti b$.  Thus, it will be
easy to eliminate backgrounds to these channels by tagging at least two
$b$ quarks that reconstruct to the mass of the $\hl$ and requiring
that the remaining jets reconstruct to the mass of the $\hl$
in the case of the $\hl\hl$ channel or to the mass of the $Z$ in the case
of the $Z\hl$ channel (lepton pairs or large missing energy with
appropriate Jacobian peak location can also be included in the latter
case). Since $b$ tagging will be possible at the NLC
with quite high efficiency and purity, we estimate that this procedure
should have an overall efficiency of 20-30\%.
With regard to the $ZZ$ channel, there is no direct irreducible background
from continuum $ZZ$ production (at tree-level), but continuum $\wp\wm$
production has a very large cross section.  The ability to separate
a $ZZ$ signal in a four-jet channel
will depend upon the precise mass resolution that can be achieved in
reconstructing two jets. To be conservative, one can require that one of
the $Z$'s decay to $\lplm$ ($l=e,\mu$); summing over
all possibilities for the second $Z$ yields a net effective branching ratio
for such channels of $\sim 0.134$.   The $\wp\wm$, and other smaller
reducible backgrounds will then be effectively eliminated by
requiring events with an $\lplm$ pair with mass near $\mz$ and,
on the opposite side, two jets or two leptons with mass near $\mz$
or else large missing energy (with appropriate Jacobian peak location).

Our results are presented in Figs.~\bslaserhh, \bslaserha,
and \bslaserhl, for the $\hh$, $\ha$ and $\hl$, respectively.
Event rates in a variety of channels are exhibited
for $\tanb=2$ and $\tanb=20$ for both $\mt=150\gev$ and $\mt=200\gev$.
We have used a uniform value of $L_{eff}\equiv F(\yh)L_{\epem}=20\fbi$
in computing event rates and have taken $\vevlam=0.8$.
Radiative corrections have been included
in the computations of $\gam\gam$ couplings and $Q\anti Q$ decays using
$\mstop=1\tev$. The $z_0=0.85$ angular cut has been included in computing
$Q\anti Q$ channel signal and background rates.
Branching ratios and efficiencies for isolating
the $\hl\hl$, $ZZ$, and $Z\hl$ final states have not been included.
We briefly summarize the conclusions that can be drawn.

We first discuss results obtained for $\mt=150\gev$ and $\tanb=2$.
Focusing on the $\hh$, we see that if most of the $ZZ$ decay events could be
employed without introducing backgrounds, then there is a region
where this mode would be preferred over the $b\anti b$ mode
which has a large irreducible background.
However, it is also clear that the best means for detecting the
$\hh$ when $\mhh<2\mt$ is in $\hh\rta \hl\hl$ decays.  The
event rate is quite large (even above $2\mt$) and, as discussed earlier,
the backgrounds are small. Above $t\bar t$ threshold,
$BR(\hh\rta t\bar t)$ is close to 1, due to the suppressed couplings
of the $\hh$ to other channels.
Detection of the $\hh$ in its $t\bar t$ decay mode from $2m_t$
up to about $500\gev$ should prove feasible.

In the case of the $\ha$, we see that the $b\anti b$ mode provides
a viable signal for $120\gev\lsim\mha\lsim 2\mt$.
However, once the background-free $Z\hl$ mode
becomes significant ($\mha\gsim 190\gev$ for $\mt=150\gev$) it provides almost
as large an absolute event rate and might prove to be the preferable mode
for detection.  In any case, there is certainly a significant range of
$\mha<2\mt$ for which $\ha$ detection should be possible in
both the $b\anti b$ and the $Z\hl$ final states, thereby allowing one to
confirm a signal for the $\ha$ seen in one channel by examining the alternative
channel. For $\mha>2\mt$, Fig.~\bslaserha\ makes it clear that the $\ha$
should be easily detectable in its $t\anti t$ decay mode.

Results for the $\hl$ are easily summarized.
Since the $\hl$ never becomes very heavy, only the $b\anti b$ final state
is relevant. The $\hl$ will be detectable in this mode for $\mhl\gsim 60\gev$.
This corresponds to the parameter space region $\mha\gsim 70\gev$.
In fact, once $\mha\gsim 2\mz$ the couplings of the $\hl$ approach their
Standard Model values, and our ability to detect the $\hl$ will
be exactly the same as our ability to find the $\hsm$ at $\mhsm\simeq\mhl$.

Let us now contrast the results obtained at $\mt=150\gev$
for the MSSM Higgs bosons in the case of $\tanb=2$ with those for
the much larger value of $\tanb=20$.
Plots corresponding to the ones given at $\tanb=2$ are presented in the second
window of Figs.~\bslaserhh-\bslaserhl.
As noted earlier, only the $b\anti b$ channel
is relevant since this is the dominant decay mode for both the $\hh$
and $\ha$ for all but small $\mha$ values (already ruled out by LEP).
In the case of the $\hh$, we see that detection in the $b\anti b$
channel should be possible for $\mhh\lsim 350\gev$, except possibly
in the region of $\mhh\sim 2\mw$.  For $\mhh$ above $350\gev$,
the $b\anti b$ event rate is not much
smaller than the $\gam\gam\rta b\anti b$ background, but the absolute
event rate is simply not adequate.  However, with four times
the luminosity assumed, detection of the $\hh$ would be possible all
the way out to $\mhh\sim 400-500\gev$. In the case of the $\ha$,
detection in the $b\anti b$ mode should be possible from the lowest
masses out to $\mha\lsim 250\gev$ at which point $\Gamma(\ha\rta\gam\gam)$
has a large dip (leading to a large dip in event rate) due to
cancellation between the $b$ and $t$ loop contributions to the
$\gam\gam$ coupling of the $\ha$.  For $\mha\gsim 350\gev$,
detection of the $\ha$ would again become possible if the luminosity
were four-fold enhanced.
Prospects for $\hl$ detection for $\tanb=20$ are very good
for all values of $\mha$. For the lowest $\mhl$ masses
(corresponding to small $\mha$),
the $b\anti b$ background is of order $10^4$ events, but the $\hl\rta
b\anti b$ signal yields about $10^3$ events, and is therefore easily
observable. (This result presumes that the
experimental resolution of $\sim 5\gev$ we have employed can be achieved.)
The enhancement of the $\hl$ cross section at small $\mhl$
when $\tanb$ is large, compared to values obtained for small $\tanb$, is due to
the fact that large $\tanb$ implies an enhanced $\hl b\anti b$ coupling.
This, in turn, leads to
an enhanced magnitude of the $b$-loop contribution to the $\gam\gam$
coupling of the $\hl$.  For
the largest values of $\mhl$ (corresponding to large $\tanb$ and large
$\mha$), $\mhl$ approaches $108\gev$. The signal, of order
500 events, is easily detectable above the smaller background, of order
80 events.

The detectability of the MSSM Higgs bosons is not greatly altered
if $\mt=200\gev$ instead of $150\gev$. Results for this choice
of $\mt$ are presented in the third and fourth windows of
Figs.~\bslaserhh-\bslaserhl.
The results are summarized simply as follows. For the $\hh$,
detection should be possible in two final state modes for nearly
all values of $\mhh$ for $\tanb=2$. At large $\tanb=20$,
the $b\anti b$ final state is the only relevant one. The signal is never
much lower than the background and detectability simply depends
on the number of accumulated events.
For the $\ha$, detection is possible in at least one, and sometimes two,
final state mode(s) for all $\ha$ masses considered, for $\tanb=2$.
Detectability of the $\ha$ when $\tanb=20$ is much more limited;
a sufficient number of signal events in the $b\anti b$ final
state is only obtained for $\mha\lsim 200\gev$.
Detecting the $\hl$ when $\mt=200\gev$ should be quite straightforward
for both moderate and large values of $\tanb$.

Thus, the prospects for MSSM Higgs boson detection at a future
$\epem$ linear collider (with center of mass energy $E_{\epem}=
500\gev$) operating in a $\gam\gam$ collider mode are promising.
Indeed, the $\gam\gam$ collider mode at the NLC
proves to be an enormously powerful tool.  For moderate $\tanb$,
detection of the $\hh$ and $\ha$ will be possible for all masses
up to about $0.8\sqrt s$, \ie\ roughly the $\gam\gam$ collider
kinematical limit, and
often in more than one final state mode. This represents a significant
increase in Higgs mass reach
as compared to a Higgs mass mass limit of roughly $0.4\sqrt s$
obtained by using the NLC in its conventional mode to search for
$\epem\rta \hh\ha$.  In the case of the lighter CP-even Higgs scalar,
detection of the
$\hl$ will be possible for $\mhl\gsim 60\gev$. The ease of detection
increases with increasing $\mt$. For large $\tanb$, $\hl$ detection
becomes possible for all $\mhl$ values, but $\hh$ and $\ha$ detection
would be confined to a low mass range (no larger than $200-300\gev$,
depending upon which Higgs is considered and the value of $\mt$).
A four-fold increase in $\gam\gam$ luminosity beyond
that assumed in this paper would allow $\hh$ detection for all masses
up to the $\egamgam$ kinematic limit when $\tanb$ is large.

In reaching these optimistic conclusions, the ability to achieve substantial
polarization for the colliding photon beams has been assumed.
In Ref.~[\ghbslaser], we have
illustrated the fact that the mass ranges for which the MSSM
Higgs bosons can be detected deteriorate significantly as
the degree of polarization decreases, especially in the
$b\anti b$ and $t\anti t$
channels. Every effort should be made to achieve as high a degree of
polarization for the colliding photons as possible.

In our analysis, we have also assumed that the supersymmetric
mass scale is sufficiently large that the Higgs bosons do not decay
to superparticle final states and charged superpartners do not significantly
contribute to the one-loop $\gam\gam$ coupling of the Higgs bosons.
Should the supersymmetric mass scale be modest in size, then the phenomenology
of the model in $\gam\gam$ collisions would be altered. In particular, once
neutralino-chargino decays of the Higgs bosons are allowed, they generally
dominate. However, we do not anticipate that it will be significantly
more difficult to detect the Higgs bosons in such
modes in $\gam\gam$ collisions.  Indeed, the neutralino-pair
channels will have no tree-level irreducible background, although one
will have to deal with backgrounds from chargino-pair events which
can be produced in continuum $\gamma\gamma$ collisions.  However,
the chargino-pair backgrounds should be no worse than the $Q\anti Q$
backgrounds discussed earlier.  Neutralino and chargino pair
events should be distinguishable
from SM-particle final states by large missing energy.
An investigation of these issues is presently in progress.

Probing the MSSM Higgs sector using back-scattered laser beams has
one other potential advantage.  Using polarized electron and
laser beams, various polarization asymmetries can be defined
that could allow an experimental determination of the CP
properties of any Higgs boson detected in $\gamma\gamma$ collisions.
\Ref\bohgun{B. Grzadkowski and J.F. Gunion, \plbj{294} (1992) 361.}
The most easily measured asymmetry is isolated by circularly
polarizing the incoming laser beams.  This asymmetry is
only non-zero if the produced
Higgs boson is a mixture of CP eigenstates. As noted earlier,
there is no CP violation in the Higgs sector of the MSSM, which
therefore predicts a null result.  Another asymmetry can be defined
using linear polarization for the initial laser beams; this latter
asymmetry would be $+1$ for production of the CP-even $\hh$ and $-1$ in the
case of $\ha$ production. Measurement of a value smaller than
unity in absolute value would contradict the MSSM.  Unfortunately,
an experimental determination of this asymmetry would require
a somewhat larger number of events than obtained for
$L_{eff}=20\fbi$ (due to the difficulty of achieving large average
linear polarizations for the back-scattered photons).  However,
such a study would clearly become of prime importance once
a Higgs boson (or two) is detected via back-scattered laser beams.

To summarize, there are three very important
motivations for an intense effort towards developing a back-scattered laser
beam facility at the NLC that has the capability for substantial
polarizations of the colliding photons.
First, such a facility provides
the only viable means for measuring the crucial $\gam\gam~Higgs$
coupling(s). Second, $\gam\gam$ collisions can greatly extend the kinematic
reach in detectable Higgs boson mass that can be achieved.
Thirdly, (polarized) $\gam\gam$ collisions provide a beautifully clean
technique
for determining the CP properties of any Higgs boson that is observed.

\vglue 0.6cm
\line{\elevenbf 7. Final Remarks \hfil}
\vglue 0.4cm

It is clear that LEP-II and the SSC combine to nearly guarantee that at least
one of the Higgs bosons of the MSSM will be discovered at one
or the other machine. If $\mt\gsim 140\gev$
(as preferred in current electroweak analyses at LEP) and $\mstop$
is of order $1\tev$, the SSC will play an important role,
allowing discovery of one or more of the MSSM Higgs over much, if not all,
of the basic $\mha$--$\tanb$ parameter space not covered by LEP-II.
Perhaps most importantly, if LEP-II discovers the light scalar Higgs, the SSC
will have a substantial chance of finding the heavy scalar, the
pseudoscalar, and/or the charged Higgs boson.  As the supersymmetric mass
scales are decreased, charginos/neutralinos
are more likely to appear in the decays of the $\ha$, $\hh$ and $\hp$ and
their non-supersymmetric-particle
SSC/LHC signals would be significantly weakened. Further study
is required to firmly establish whether Higgs detection in
the chargino/neutralino modes would be feasible in such a case,
but preliminary studies allow for some optimism. Of course,
as the supersymmetric mass scale decreases,
radiative corrections to $\mhl$ would be relatively smaller, and the
probability of discovering the $\hl$ at LEP-II is substantially increased.
In general, for a light top quark and/or light squarks,
LEP-II will play the most important role, and sensitivity of the SSC to the
MSSM Higgs bosons could be small. For this reason it is critical
that the energy achieved by LEP-II be maximized.

Of course, we have also seen that a sufficiently energetic linear $\epem$
collider with adequate luminosity would provide a fairly ideal machine for the
detection of the MSSM Higgs bosons. Operating in the
conventional $\epem$ collision mode, detection of the $\hl$ would be
certain and the $\ha$, $\hh$ and $\hp$ could be discovered for
masses up to $\mha\sim\mhh\sim\mhp\sim 0.4 \sqrt s$.

However, there seems to be a significant possibility that only the
$\hl$ would be observed with the combination of an NLC and the SSC/LHC.
Indeed, GUT scenarios suggest that SUSY mass scales
(in particular $\mstop$ and the ino masses) are not very large,
but that $\mha\sim\mhh\sim\mhp\gsim 200\gev$.  In this case,
discovery of the $\ha$, $\hh$, and $\hp$ at SSC/LHC could prove
quite difficult, while the conventional $\epem\rta\ha\hh$ and
$\epem\rta\hp\hm$ pair production processes would be
either very close to or beyond the kinematic limit
of a $\sqrt s =500\gev$ NLC and could not be used to probe
masses much beyond $\sim 0.4\sqrt s=200\gev$. Only a back-scattered laser beam
facility at the NLC provides a clear-cut possibility of probing
the higher $\ha,\hh,\hp$ masses found in many GUT scenarios.
With good polarization for the colliding photon beams, $\ha$ and $\hh$
masses up to $\sim 0.8\sqrt s$ can be probed for moderate
$\tanb$ at easily achieved effective $\gam\gam$ luminosities.
(Full coverage of this mass range for large $\tanb$ requires about four times
as much luminosity.)
The roughly $400\gev$ upper mass reach that could be achieved encompasses
many of the GUT scenario predictions for $\ha$ and $\hh$ masses.

Overall, the various accelerators of the next decade will
almost certainly unlock the secrets of a supersymmetric
electroweak symmetry breaking sector, providing many possibilities
for relatively detailed verification of its features.
In conjunction with the probable discovery of supersymmetric partners,
such as the gluino and squark, detailed consistency checks
of the predictions (including radiative corrections) for masses and couplings
of the Higgs bosons will become possible.

\vglue 0.6cm
\line{\elevenbf 8. Acknowledgements\hfil}
\vglue 0.4cm
Some of the work described in this review originated at the U.C. Davis
Workshop on Higgs/EWSB Physics at Hadron Super Colliders.
Contributions from my collaborators, H.E. Haber and L.H. Orr,
are gratefully acknowledged. This work was supported in part by the Department
of Energy.

\refout

\end